\documentclass[12pt]{article}
\usepackage{amssymb,amsmath,amscd}
\makeatletter

\@addtoreset{equation}{section}
\def\section{\@startsection {section}{1}{\z@}{-2.25ex plus -1ex minus
 -.2ex}{1.0ex plus .2ex}{\large\bf}}
\def\subsection{\@startsection{subsection}{2}{\z@}{-2.0ex plus%
 -1ex minus -.2ex}{0.5ex plus .2ex}{\bf}}

\def\Ad{\mbox{Ad}}

\def\ba{{\mbox{\boldmath $a$}}}
\def\bx{{\mbox{\boldmath $x$}}}
\def\bs{{\mbox{\boldmath $s$}}}

\def\bj{{\mbox{\boldmath $j$}}}

\newcommand{\gothf}{\mathcal F }
\newcommand{\gothg}{\mathfrak g }

\newcommand{\CC}{\mathbb{C}}

\def\bea{\begin{eqnarray}}
\def\eea{\end{eqnarray}}
\newtheorem{theorem}{Theorem}[section]

\newtheorem{lemma}[theorem]{Lemma}

\newtheorem{definition}[theorem]{Definition}

\def\bmz{\left(\begin{array}{2,2}}
\def\emz{\end{array}\right)}
\def\bmd{\left(\begin{array}{3,3}}
\def\emd{\end{array}\right)}

\newcommand{\inv}[0]{{-1}}

\newcommand{\mi}[0]{{M_i}}
\newcommand{\ai}[0]{{A_i}}
\newcommand{\bi}[0]{{B_i}}

\newcommand{\aj}[0]{{A_j}}
\newcommand{\bjj}[0]{{B_j}}

\newcommand{\me}[0]{{M_1}}
\newcommand{\aee}[0]{{A_1}}

\newcommand{\mf}[0]{{M_n}}

\newcommand{\bff}[0]{{B_g}}

\newcommand{\cif}[0]{\mathcal{C}^\infty}

\newcommand{\prgr}{G\ltimes \mathfrak{g}^*}

\newcommand{\repind}[0]{{\mu_1s_1\ldots\mu_ns_n}}

\newcommand{\call}[0]{\mathcal{L}}

\newcommand{\calc}[0]{\mathcal{C}}

\newcommand{\flh}[3] {f_{{#1}{#2}}^{\;\;\;\;{#3}}\;}

\newcommand{\tenltimes}[0] {\mbox{$\subset\!\!\!\!\!\!\times$}}
\newcommand{\Gh}[0]{G^{n+2g}}

\newcommand{\ntg}[0]{{n+2g}}

\newcommand{\mapcl}[0]{\text{Map}(S_{g,n})}
\newcommand{\pmapcl}[0]{\text{PMap}(S_{g,n})}
\newcommand{\mindee}{\mbox{$\!\setminus D$}}
\newcommand{\mapcld}[0]{\text{Map}(S_{g,n}\mindee )}
\newcommand{\pmapcld}[0]{\text{PMap}(S_{g,n}\mindee )}

\begin{document}
\parskip 6pt
\parindent 0pt
\begin{flushright}
HWM-03-32\\
EMPG-03-23\\
hep-th/0312049
\end{flushright}

\begin{center}
\baselineskip 24 pt
{\Large \bf  Mapping class group actions in Chern-Simons theory with
  gauge group $\prgr$}

\vspace{1cm}
{\large C.~Meusburger\footnote{\tt  cm@ma.hw.ac.uk}
and    B.~J.~Schroers\footnote{\tt bernd@ma.hw.ac.uk} \\
Department of Mathematics, Heriot-Watt University \\
Edinburgh EH14 4AS, United Kingdom } \\

\vspace{0.5cm}

{  15 Feburary 2004}

\end{center}

\begin{abstract}
\baselineskip 12pt
\noindent  We study the action of the mapping class group of an
oriented genus $g$ surface
with $n$ punctures and a
disc removed  on a
Poisson algebra which arises in the
combinatorial description of Chern-Simons gauge theory 
 when the gauge group is a semidirect product
$\prgr$ . We prove that the mapping class group acts on this algebra via Poisson isomorphisms and express the action of
Dehn twists in terms of an infinitesimally generated $G$-action. We
 construct a mapping class group representation 
  on the representation spaces of the
associated quantum algebra and show that Dehn twists can be implemented
 via the ribbon element of the quantum double $D(G)$ and 
 the exchange of punctures
 via its universal $R$-matrix.
\end{abstract}


\section{Introduction}

This paper investigates  a  mapping class group action
 on a certain
Poisson algebra and on the representation spaces of the associated
quantum algebra. This Poisson algebra, in the following referred to as
flower algebra, plays an important role in the combinatorial
description of the phase space of Chern-Simons theory with semidirect
product gauge groups of the form $\prgr$, where $G$ is a Lie group,
$\gothg^*$ the dual of its Lie algebra and $G$ acts on $\gothg^*$
in the coadjoint representation. Its classical structure and
quantisation were studied in an earlier paper \cite{we2}.

Although mapping class groups and Chern-Simons gauge theories are 
research topics in their own
right, our interest in them is motivated by their relevance to
physics. Chern-Simons gauge theory with gauge group $\prgr$ occurs in
the Chern-Simons formulation of (2+1)-dimensional gravity with
vanishing cosmological constant \cite{Witten1},
where, depending on the signature of the spacetime, the gauge group is
the three dimensional Poincar\'e or Euclidean group. For spacetimes of
topology $\mathbb{R}\times S$, where $S$ is an oriented
surface of arbitrary genus, possibly with punctures and boundary components,
  large diffeomorphisms of the surface $S$ give rise to
Poisson symmetries or canonical transformations on the phase space
\cite{we}. There is evidence suggesting that these symmetries play an
important role in the physical interpretation of the theory, in 
particular for the dynamics of gravitationally
interacting particles \cite{Carlip, BMS}.

The flower algebra emerges in a description of the moduli space of
flat connections on a genus $g$ surface  $S_{g,n}$ with $n$ punctures,
 discovered by Fock and Rosly
\cite{FR} and developed further by Alekseev, Grosse and Schomerus
\cite{AGSI,AGSII,AS}. In this description, the moduli space is given
as a quotient of the space of holonomies  associated to a set of 
$n+2g$ generators of
the surface's fundamental group equipped with a certain Poisson
structure. In our case, these
holonomies are elements of $\prgr$, and one obtains a Poisson structure
on the manifold $(\prgr)^\ntg$. The flower algebra  is the algebra of
 a particular class of functions on $(\prgr)^\ntg$ with this Poisson bracket. 
In \cite{we2}, we
investigated the classical
properties of this Poisson algebra and constructed the corresponding
quantum algebra and its irreducible Hilbert space representations.
In this paper, we show that the mapping class group $\mapcld$ of
the surface $S_{g,n}$ with a disc $D$ removed acts on the flower algebra and
determine the associated quantum action.

We prove that the mapping class group  action on the flower algebra is a
Poisson action and show  that the action of Dehn twists around embedded 
curves on the surface $S_{g,n}\mindee$ can be expressed in terms of  
a $G$-action that is
infinitesimally generated via the Poisson bracket. For the case where
the exponential map $\exp:\gothg\rightarrow G$ is surjective, we
 give explicit Hamiltonians whose flow by one unit is 
equal to the action of Dehn twists. These flows 
are special examples of the Hamiltonian ``twist flows'' 
 studied by Goldman in \cite{Goldman0}, but the particular 
Hamiltonians we consider and their relation with Dehn 
twists appear to be new.

We then demonstrate how these classical features are mirrored by
corresponding structures in the quantum theory. We show that elements
of the mapping class group act as algebra automorphisms on the
quantised flower algebra and implement this action on its
representation spaces. This allows us to relate the quantum action of
the mapping class group to different representations of the quantum double
$D(G)$ of the group $G$. We find a canonical way  of associating a
representation of the quantum double $D(G)$ to each embedded curve, such that the action of the corresponding Dehn twist
is given by the ribbon element of $D(G)$. We find an implementation of 
the  exchange of punctures
on the surface $S_{g,n} \mindee$  in terms of the action
of the universal $R$-matrix in the tensor product of two representations
 of $D(G)$, familiar from the theory of quantum groups. 
 
The paper is structured as follows: Sect.~2 gives a summary of our
results in \cite{we2} required for the understanding of this article,
which is necessarily rather condensed. In Sect.~3 we discuss the
mapping class group action on the classical flower algebra and express
the action of Dehn twists in terms of infinitesimally generated group actions
as outlined above. Sect.~4 investigates the corresponding
quantum action and relates it to representations of the quantum double
$D(G)$, followed by a brief outlook in Sect.~5. The appendix lists a 
set of generators of the mapping class
group and their actions on the fundamental group 
$\pi_1(S_{g,n}\mindee)$.


\section{The phase space of Chern-Simons gauge theory with gauge group $\prgr$ and the flower algebra}

\subsection{Notation and conventions}
We consider groups $\prgr$ which are the
semidirect product of a finite-dimensional, simply connected and connected
 Lie group $G$ and the dual $\gothg^*$ of its  Lie algebra
$\gothg=\text{Lie}\,G$, viewed as an abelian group. 
All Lie algebras are vector spaces over
$\mathbb{R}$ unless stated otherwise, and Einstein summation convention is used throughout the paper. Following the conventions of \cite{MarsRat}, we define $\Ad^*(g)$ to be the algebraic dual of $\Ad(g)$, i.~e.~
\bea
\langle \Ad^*(g)\bj, \xi\rangle=\langle \bj, \Ad(g)\xi \rangle\qquad\forall \bj\in\gothg^*, \xi\in\gothg, g\in G,
\eea
so that the coadjoint action of $g\in G$ is given by $\Ad^*(g^\inv)$.
Writing elements of $\prgr$ as $(u,\ba)$ with $u\in G$ and
$\ba\in\gothg^*$, we have the group multiplication law
\bea
\label{groupmult}
(u_1,\ba_1)\cdot(u_2,\ba_2)=(u_1\cdot u_2,\ba_1+\Ad^*(u_1^\inv)\ba_2).
\eea
We also use the parametrisation
\bea
\label{gparam}
(u,\ba)=(u,-\Ad^*(u^\inv)\bj)\qquad\text{with}\; u\in G,\;\ba,\bj\in\gothg^*,
\eea
where, as explained in \cite{we2}, the pair $(u,\bj)$ should be thought of as
an element of the dual Poisson-Lie group.

Let  $J_a$, $P^a$, $a=1,\ldots,\text{dim}\,G$, denote the generators
of the Lie algebra $\text{Lie}\, (\prgr)=\gothg\oplus\gothg^*$, such
that the generators $J_a$ form a basis of
$\gothg=\text{Lie}\,G$ and the generators $P^a$ a basis of
$\gothg^*$. Then the commutator of the Lie algebra
$\gothg\oplus\gothg^*$ is given by  
\bea
\label{commutator}
[J_a,J_b]=\flh a b c J_c\qquad[J_a,P^b]=-\flh a c b P^ c\qquad[P^a,P^b]=0,
\eea
where $\flh a b c$ are
the structure constants of $\gothg$.
We denote by  $J_a^R$, $J_a^L$ the  left-and right-invariant vector fields on
$G$ associated to the generators $J_a$
\begin{align}
\label{gvecfields}
&J_a^R F(u):=\frac{\text{d}}{\text{d}t}|_{t=0}F(u e^{tJ_a})\qquad
J_a^L F(u):=\frac{\text{d}}{\text{d}t}|_{t=0}F(e^{-tJ_a}u) & &\forall
u\in G,\, F\in\cif(G).
\end{align}


\subsection{The flower algebra for gauge group $\prgr$}

The flower algebra plays an important role in the description
of the phase space of Chern-Simons theory with semidirect product
gauge groups of type $\prgr$. Mathematically, this phase space is the
moduli space of flat $\prgr$-connections on the surface $S_{g,n}$, and
it can be described as a quotient of the space of holonomies
 holonomies associated to a set of generators of the
surface's fundamental group. Fock and Rosly \cite{FR} and Alekseev,
Grosse and Schomerus \cite{AGSI,AGSII,AS} defined a Poisson structure
on the space of holonomies, which, via Poisson reduction, gives rise
to the canonical Poisson structure on the moduli space \cite{Goldman, AB}.
For Chern-Simons theory with compact, semisimple gauge groups, this
Poisson structure on the space of holonomies was investigated by
Alekseev, Grosse and Schomerus \cite{AGSI,AGSII,AS} and quantised via
their formalism of combinatorial quantisation of Chern-Simons gauge
theories. The case of (non-compact and non-semisimple) semidirect product gauge groups of type $\prgr$
was studied in \cite{we2}, where we discussed the properties of this Poisson structure and developed a quantisation procedure.

In order to define the flower algebra for Chern-Simons theory with 
gauge group $\prgr$ on a punctured surface, we need a set of 
generators of the surface's
fundamental group. Both the  fundamental group $\pi_1(S_{g,n})$ of a genus
$g$ surface $S_{g,n}$ with $n$ punctures 
and the fundamental group of the associated 
surface $S_{g,n}\mindee$ with 
a disc $D$ removed 
is generated by 
the equivalence classes
 loops $m_i$, $i=1,\ldots,n$, around  the  punctures and two curves
$a_j$, $b_j$, $j=1,\ldots,g$, for each handle, shown in  Fig.~1.  
In the case of the surface $S_{g,n}\mindee$ with a disc
removed they generate the fundamental group freely, whereas for the 
surface  $S_{g,n}$ they are 
subject to the relation
 \bea
\label{pirel}
[b_g,a_g^{-1}]\cdot\ldots\cdot [b_1,a_1^{-1}]\cdot m_n\cdot\ldots\cdot m_1=1,
\quad
\mbox{with} \quad[b_i,a_i^\inv]=b_i a_i^\inv b_i^\inv a_i.
\eea
\vbox{
\vskip .3in
\input epsf
\epsfxsize=12truecm
\centerline{
\epsfbox{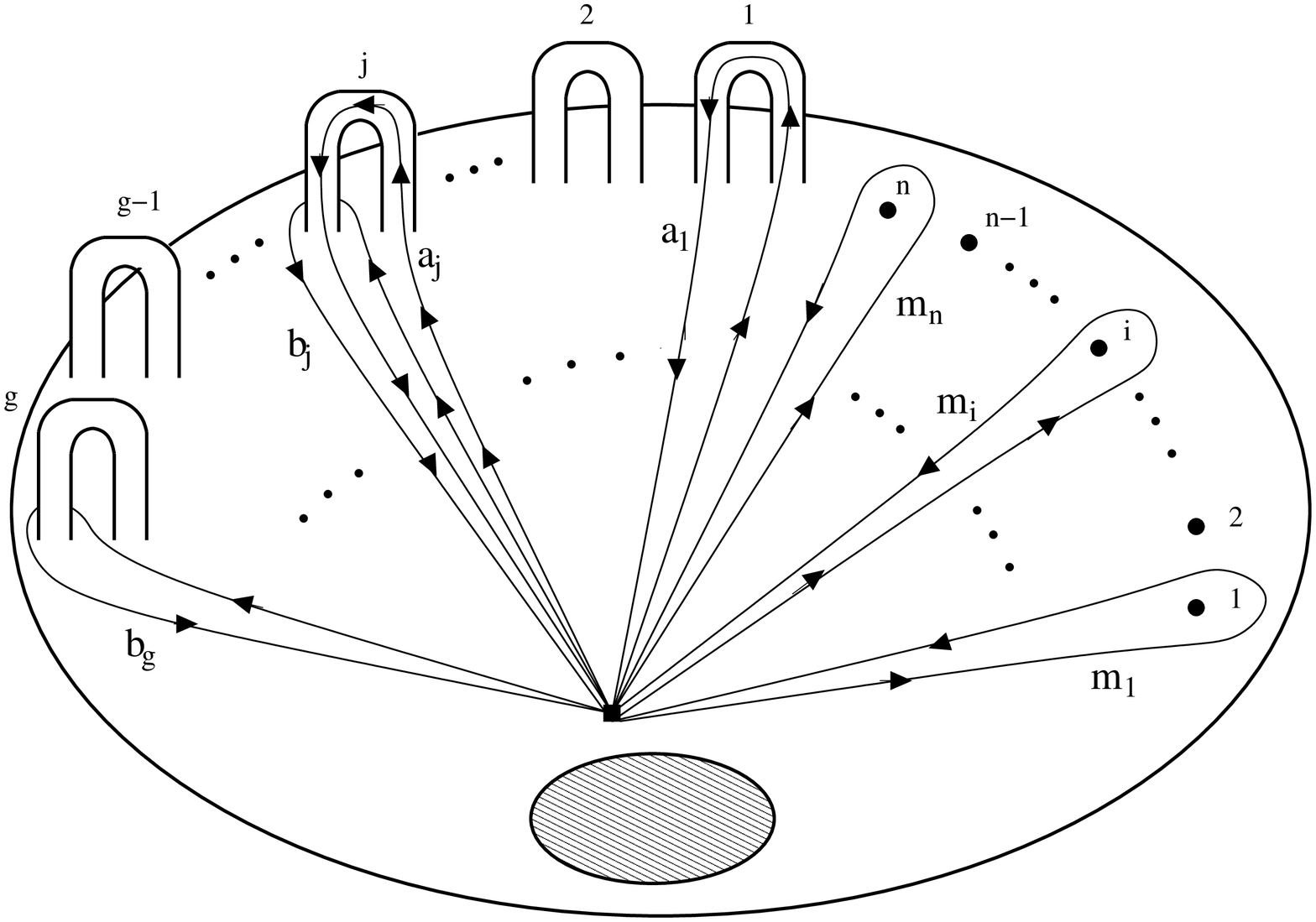}
}
\bigskip
{
\centerline{\bf Fig.~1 }
}
\centerline{\footnotesize The generators of the
fundamental group of the surface $S_{g,n}\mindee $}
\bigskip}

In the rest of the paper, we do not distinguish notationally between
closed curves on $S_{g,n}$  and $S_{g,n}\mindee$,
and their equivalence classes in the
fundamental group $\pi_1(S_{g,n})$ and $\pi_1(S_{g,n}\mindee)$.

Whereas the holonomies $A_j=\text{Hol}(a_j)$,
$B_j=\text{Hol}(b_j)$ associated to
each handle are general elements of the group $\prgr$, the
holonomies $M_i=\text{Hol}(m_i)$ around the punctures lie in
fixed $\prgr$-conjugacy classes 
\bea
\label{conjugcla}
\calc_{\mu_is_i}=\{ (v,\bx)\cdot (g_{\mu_i},-\bs_i)\cdot
(v,\bx)^\inv\;|\; (v,\bx)\in \prgr\}.
\eea
For a geometrical interpretation of the labels $\mu_i$ and $s_i$ we
refer the reader to \cite{we2}.

By applying the work of Fock and Rosly \cite{FR} and Alekseev, Grosse
and Schomerus \cite{AGSI,AGSII, AS} to the case of gauge group
$\prgr$, one obtains a Poisson structure on $(\prgr)^\ntg$. However, for gauge groups $\prgr$ it is more
convenient to work with the slightly different formulation used in \cite{we2}.
We parametrise the holonomies according to \eqref{gparam} as
$X=\text{Hol}(x)=(u_X,-\Ad^*(u_X^\inv)\bj^X)$ for
$X\in\{M_1,\ldots,M_n,A_1,B_1,\ldots,A_g,B_g\}$, expand the vectors
$\bj^X$ as  $\bj^X=j^X_bP^b$ and denote by the same symbol  the coordinate functions 
\bea
\label{jmap}
j^X_a\in\cif((\prgr)^{n+2g}): (M_1,\ldots,M_n,A_1,B_1,\ldots,A_g,B_g)\mapsto
j^X_a.
\eea Instead of the algebra  $\cif((\prgr)^{n+2g})$ we then consider the algebra
generated  by the functions
in $\cif(\Gh)$ together with these maps $j^X_a$ with the Poisson
structure given below.

\begin{definition} (Flower algebra for groups $\prgr$)
\label{flower}

The flower algebra $\gothf$ for gauge group $\prgr$ on a genus $g$
 surface $S_{g,n}$ with $n$ punctures is the commutative Poisson algebra
\bea
\label{flowerdef}
\gothf=S\left(\bigoplus_{k=1}^{n+2g}\gothg\right)\otimes\cif(G^{n+2g}),
\eea
where $S\left(\bigoplus_{k=1}^{n+2g}\gothg\right)$ is the symmetric envelope of
the real Lie algebra $\bigoplus_{k=1}^{n+2g}\gothg$, i.e. the polynomials with real
coefficients on the vector space $\bigoplus_{l=1}^{n+2g}\gothg^*$. In
terms of a fixed basis
$\mathcal{B}=\{j_a^\mi,j_a^{A_k},j_a^{B_k},\; $i=1,\ldots,n,\; k=1,\ldots,g,\;
a=1,\ldots,\text{dim}\,G\},  its Poisson structure is given by
\begin{align}
\label{poiss}
&\{j_a^X \otimes 1,j_b^X\otimes 1\}=-\flh abc j^X_c\otimes
1\nonumber\\ 
&\{j_a^X\otimes 1,j_b^Y\otimes 1\}=- \flh d b c
j_c^Y\otimes(\delta_a^{\;\;d}-\Ad^*(u_X)_a^{\;\;d})\qquad\qquad\forall
X,Y\in\{M_1,\ldots,B_g\},\, X<Y\nonumber\\
&\{j^\ai_a\otimes 1,j^\bi_b\otimes 1\}=-\flh a b c j^\bi_c\otimes 1\;\qquad\qquad\qquad\qquad\qquad\forall i=1,\ldots,g\nonumber\\
\nonumber\\ 
&\{j^\mi_a\otimes 1, 1\otimes F\}=-1\otimes
(J_a^{R_\mi}+J_a^{L_\mi})F-1\otimes
(\delta_a^{\;\;b}-\Ad^*(u_\mi)_a^{\;\;b})\left(\sum_{Y>\mi}
  (J_b^{R_Y}+J_b^{L_Y})F\right)\nonumber\\ 
   \intertext{}  
&\{j^\ai_a\otimes 1, 1\otimes F\}=-1 \otimes (J^{R_\ai}_a+J^{L_\ai}_a)F-1\otimes (J_a^{R_\bi}+J_a^{L_\bi})F-1\otimes \Ad^*(u_\bi^\inv u_\ai)_a^{\;\;b}J_b^{R_\bi}\nonumber\\
&\qquad\qquad\qquad\qquad-1\otimes
(\delta_a^{\;\;b}-\Ad^*(u_\ai)_{a}^{\;\;b})\left(\sum_{Y>\ai}
  (J_b^{R_Y}+J_b^{L_Y})F\right)\nonumber\\   
&\{j^\bi_a\otimes 1, 1\otimes F\}=-1\otimes J_a^{L^\ai} F-1\otimes (J_a^{R_\bi}+J_a^{L_\bi})F\nonumber\\
&\qquad\qquad\qquad\qquad-1\otimes (\delta_a^{\;\;b}-\Ad^*(u_\bi)_a^{\;\;b})\left(\sum_{Y>\bi} (J_b^{R_Y}+J_b^{L_Y})F\right),
\end{align}
where $F\in\cif(G^{n+2g})$, $M_1<\ldots<M_n<A_1,B_1<\ldots<A_g,B_g$
and $J_a^{L_X}$, $J_a^{R_X}$ denote the right- and left invariant vector
fields \eqref{gvecfields} on the different copies of $G$.
\end{definition}
Note that this definition does not restrict the holonomies $M_i$
associated to the punctures to fixed $\prgr$-conjugacy classes
$\calc_{\mu_i s_i}$. Instead, these conjugacy classes arise as the
symplectic leaves of the Poisson structure \eqref{poiss} on $(\prgr)^\ntg$. Furthermore,
we showed in \cite{we2} by extending the work of Alekseev and Malkin \cite{AMII}
to groups of type $\prgr$
that the Poisson structure on the symplectic leaves is given by a
symplectic potential $\Theta$ on $(\prgr)^\ntg$. This potential can be
expressed in terms of the holonomies as follows

\begin{theorem} (Symplectic leaves and decoupling)

\label{symp}
The symplectic leaves of the Poisson manifold $(\prgr)^\ntg$ with bracket \eqref{poiss} are
  of the form
  $\calc_{\mu_1s_1}\times\ldots\times\calc_{\mu_ns_n}\times T^*(G)^{2g}$,
  where  $\calc_{\mu_is_i}$ denote $\prgr$-conjugacy classes as in
  \eqref{conjugcla}.   Let $\omega_\gothf$ denote the symplectic form on these
  symplectic leaves, define a map $\beta:G^\ntg\rightarrow G^\ntg$
\begin{align}
\label{betadef}
\beta:(v_{M_1},\ldots, v_{M_n},u_\aee,\ldots,u_\bff)\mapsto &(v_{M_1}g_{\mu_1}
v_{M_1}^\inv, \ldots,v_{M_n}g_{\mu_n}
v_{M_n}^\inv, u_{A_1},\ldots, u_{B_g})\\
&=:(\beta_{M_1}(v_{M_1},\ldots,,u_\bff),
\ldots,\beta_{B_g}(v_{M_1},\ldots,u_\bff))\nonumber
\end{align} and extend it trivially to a map
$\tau:(\prgr)^{n+2g}\rightarrow (\prgr)^\ntg$ via
\begin{align}
\label{taudef}
\tau:
&(v_{M_1},\bj^{M_1},\ldots,v_{M_n},\bj^{M_n},u_\aee,\bj^{\aee},\ldots,u_\bff,\bj^{\bff})\mapsto\\&\qquad\qquad\qquad\qquad\qquad(\beta_{M_1}(v_{M_1},\ldots,u_\bff),\bj^{M_1},\ldots,\beta_{B_g}(v_{M_1},\ldots,u_\bff),\bj^\bff).\nonumber
\end{align}
Then, the pull-back $\tau^*\omega_\gothf$ of $\omega_\gothf$ with $\tau$ coincides with the exterior derivative of the
 symplectic potential 

\begin{align}
\label{thetaundec}
\Theta=&\sum_{i=1}^n \langle\; d(u_{M_{i-1}}\cdots u_{M_1})(u_{M_{i-1}}\cdots u_{M_1})^\inv- dv_\mi v_\mi^{-1}\,,\, j^\mi_a
P^a\rangle\\
+&\sum_{i=1}^g\langle\; d(u_{K_{i-1}}\cdots u_{M_1})(u_{K_{i-1}}\cdots u_{M_1})^\inv\,,\, j_a^\ai P^a\rangle\nonumber\\
&\qquad-\langle\; d(u_\ai^\inv u_\bi^\inv u_\ai u_{K_{i-1}}\cdots u_{M_1})(u_\ai^\inv u_\bi^\inv u_\ai u_{K_{i-1}}\cdots u_{M_1})^\inv\,,\,
j_a^\ai P^a\rangle\nonumber\\
+&\sum_{i=1}^g\langle\; d(u_\ai^\inv u_\bi^\inv u_\ai u_{K_{i-1}}\cdots u_{M_1})(u_\ai^\inv u_\bi^\inv u_\ai u_{K_{i-1}}\cdots u_{M_1})^\inv\,,\,
j_a^\bi P^a\rangle\nonumber\\
&\qquad-\langle\; d( u_\bi^\inv u_\ai u_{K_{i-1}}\cdots u_{M_1})( u_\bi^\inv u_\ai u_{K_{i-1}}\cdots u_{M_1})^\inv\,,\,
j_a^\bi P^a\rangle\nonumber,
\end{align}
where $u_{K_i}=[u_{B_i}, u_{A_i}^\inv]=u_\bi u_\ai^\inv u_\bi^\inv u_\ai$.
\end{theorem}

{\bf Proof:} This follows from Theorems 2.4., 2.5. in \cite{we2} by expressing the symplectic form $\Theta$ defined there in terms of the the coordinate functions $j^X_a$ \eqref{jmap}.
\hfill $\Box$

Under the pull-back $\tau^*$ the conjugation action on the group
elements associated to the punctures gets mapped to
left-multiplication. Hence,  if we consider the Poisson algebra generated by
the maps $j^X_a$ in \eqref{jmap} and functions in $\cif(G^\ntg)$ with
the bracket induced by the symplectic potential $\Theta$
\eqref{thetaundec} we obtain a modified bracket $\{\,,\,\}_\Theta$ on
$S\left(\bigoplus_{k=1}^{n+2g}\gothg\right)\otimes\cif(G^{n+2g})$ that
is given by \eqref{poiss} with the exception of
\begin{align}
\label{poisstheta}
\{j^\mi_a\otimes 1, 1\otimes F\}_\Theta=&-1\otimes J_a^{L_\mi} F-1\otimes (\delta_a^{\;\;b}-\Ad^*(u_\mi)_a^{\;\;b})\left(\sum_{j=i+1}^n J^{L_{M_j}}_b F\right) \\
&-1\otimes (\delta_a^{\;\;b}-\Ad^*(u_\mi)_a^{\;\;b})\left( \sum_{j=1}^g (J_b^{R_{A_j}}+J_b^{L_{A_j}}+J_b^{R_{B_j}}+J_b^{L_{B_j}})F\right).\nonumber
\end{align}

In \cite{we2}, we made use of this link between the flower algebra Poisson structure \eqref{poiss}
and the symplectic potential  $\Theta$ \eqref{thetaundec} to construct
the corresponding quantum algebra and to investigate its
representation theory. We obtained the following theorem
\begin{theorem}(Quantum flower algebra)
\label{qflower}

 The quantum algebra for the flower algebra in Def.~\ref{flowerdef}
  is the associative algebra
\bea
\label{flowernew}
\hat \gothf=U\left(\bigoplus_{k=1}^{n+2g} \gothg\right)
\tenltimes\cif(G^{n+2g},\CC),
\eea
with the multiplication defined by
\begin{align}
&(\xi\otimes F)\cdot(\eta\otimes K)=\xi\cdot_U \eta\otimes FK+i \hbar\;
  \eta\otimes F\{\xi\otimes 1,1\otimes K\},
\end{align}
where $\xi,\eta\in \bigoplus_{k=1}^{n+2g}\gothg, F,K\in\cif(G^{n+2g},\CC)$ and  $\cdot_U$ denotes the multiplication in
$U\left(\bigoplus_{k=1}^{n+2g}\gothg\right)$. The bracket $\{\,,\,\}$ is given by \eqref{poiss}.
\end{theorem}
We found in \cite{we2} that the representation theory of this algebra is best investigated in the framework of representation theory of transformation group algebras. As the discussion is quite technical, we summarise only the main result and refer the reader to \cite{we2} for further details and some technical assumptions on the group $G$. Further information about transformation group algebras can be found in \cite{kM}, which gives a treatment closely related to our situation.
\begin{theorem}(Representations of the quantum flower algebra)
\label{flowrep}

 Under the technical assumptions on the group $G$ in \cite{we2},
 the irreducible representations of the flower algebra are labelled by $n$
  $G$-conjugacy classes $\calc_{\mu_i}=\{g g_{\mu_i}g^\inv| g\in G\}$, $i=1,\ldots,n$, and
 irreducible unitary Hilbert space
  representations $\Pi_{s_i}: N_{\mu_i}\rightarrow V_{s_i}$
 of the  stabilisers $N_{\mu_i}=\{g\in G| g g_{\mu_i} g^\inv=g_{\mu_i} \}$ of
  chosen elements $g_{\mu_1},\ldots g_{\mu_n}$ in those conjugacy
  classes.
Consider the space
\begin{align}
&L^2_{\mu_1s_1\ldots\mu_n s_n}=\big\{\psi: G^{n+2g} \rightarrow V_{s_1}
\otimes\ldots\otimes V_{
 s_n})\;|\;\psi( v_1 h_1,\ldots,v_{M_n} h_n,  u_{A_1},\ldots, u_{B_g})\nonumber\\
&\quad=(\Pi_{s_1}(h_1^\inv)\otimes\ldots\otimes
  \Pi_{s_n}(h_n^\inv))\;\psi(v_{M_1},\ldots,
  v_{M_n},u_{A_1},\ldots,u_{B_g})\; \forall h_i\in    N _{\mu_i}\; and
  ||\psi||^2 <\infty\big\},\nonumber
\end{align}
with inner product
\begin{align}
\langle \psi,\phi\rangle=\int_{G/N_{\mu_1}\times\ldots\times G/N_{\mu_n}\times G^{2g} } &\left(\,\psi\,,\, \phi\,\right)
(v_{M_1},\ldots,v_{M_n},u_{A_1},\ldots u_{B_g})\\
&\qquad\qquad dm_1(v_{M_1}\cdot N_1)\cdots
dm_n({v_{M_n}\cdot N_n})\cdot du_{A_1}\cdots du_{B_g},\nonumber
\end{align}
where  $(\,,\,)$  is the canonical inner product on the tensor
product of Hilbert spaces $V_{s_1}\otimes \ldots\otimes V_{s_n}$.
Then the representation spaces $V_\repind$ are obtained from $L^2_\repind$ by dividing out the zero-norm states.
 The quantum flower algebra acts on a dense subspace
 $V_\repind^\infty$ of $C^\infty$-vectors \cite{we2}
according to
\begin{align}
\label{quantact}
&\Pi_\repind(X\otimes F)\psi=-i\hbar (F\circ\beta)\cdot \{X,\psi\}_\Theta\\
&\Pi_\repind(1\otimes F)\psi=(F\circ\beta)\cdot\psi\nonumber
\end{align}
where $F\in\cif(G^\ntg)$, $X\in\gothg^\ntg$ and   $\beta$ is given by \eqref{betadef}.
\end{theorem}


\section{The classical action of the mapping class group}
\label{classmapact}

\subsection{ Poisson action of the mapping class group}
As a diffeomorphism invariant theory, Chern-Simons theory on
$\mathbb{R}\times S_{g,n}$ is in particular invariant under 
orientation preserving  diffeomorphisms
of $S_{g,n}$ which are not connected to the identity.
The equivalence classes of such diffeomorphisms  constitute the surface's
mapping class group $\text{Map}(S_{g,n})$ \cite{birmorange}. 
Therefore, one would expect
the mapping class group to act via Poisson isormophisms on the
phase space of Chern-Simons theory, the moduli space of flat
connections. The moduli space is obtained from the set of holonomies
around the curves in Fig.~1 by imposing the relation \eqref{pirel} 
and dividing by conjugation. In the flower algebra, by contrast,  the 
relation \eqref{pirel} is not imposed
and therefore there is no group action of $\mapcl$ on the flower algebra.
We shall now show that, instead,  there is a a group action of 
the mapping class group $\mapcld$ on the flower algebra, and 
that this action is Poisson. 

The mapping class group $\mapcld$ is the group of equivalence
classes of orientation preserving 
diffeomorphisms of $ S_{g,n}\mindee$ which 
fix the punctures as a set
and  the boundary of the disc $D$ pointwise; diffeomorphisms 
are equivalent if they differ by one which is isotopic to the 
identity.
It contains elements that leave the punctures invariant as 
well as elements that exchange different punctures. The former, 
by definition, form the pure mapping class group 
$\pmapcld$ related to the mapping class group by the 
short exact sequence
\bea
\label{sequen}
1\rightarrow\pmapcld\xrightarrow{i}
\mapcld \xrightarrow{\pi}
S_n \rightarrow 1,
\eea
where $i$ is the canonical  embedding and $\pi:\mapcld\rightarrow S_n $
 the projection onto the
symmetric group that assigns to each element of the mapping 
class group the associated permutation of the punctures. As explained in 
\cite{birmgreen,birmorange}, the pure mapping class group 
$\pmapcld$ is generated by Dehn twists around a 
set of embedded curves, and 
a set of generators of the full mapping class group 
$\mapcl$ can be obtained by supplementing this set with 
 $n-1$ elements
  which get mapped to the elementary
transpositions
via $\pi$. A set of generators of the pure and full
mapping class group and their action on the fundamental group is 
given in in the appendix.

The action of the mapping class group on the flower algebra arises in
the following way. Elements $\lambda\in\mapcld$ act as automorphisms on
the fundamental group $\pi_1(S_{g,n}\mindee)$ and give rise to
transformations of the holonomies along the generating curves $m_i$,
$a_j$, $b_j$, thus inducing a map $(\prgr)^\ntg\rightarrow (\prgr)^\ntg$ 
which we denote by $\Lambda$. Explicitly, we have
\begin{align}
\label{grmapdef}
\Lambda: (\text{Hol}(m_1),\ldots,\text{Hol}(m_n),\text{Hol}(a_1),
\text{Hol}(b_1),\ldots, \text{Hol}(a_g),&\text{Hol}(b_g))\\
&\mapsto (\text{Hol}(\lambda(m_1)),\ldots,\text{Hol}(\lambda(b_g))).\nonumber
\end{align}
The push-forward with $\Lambda$ defines a map $\Lambda_*: 
\cif((\prgr)^\ntg)\rightarrow \cif((\prgr)^\ntg)$, which maps the 
flower algebra into itself. We write $\Lambda_G$ for the restriction 
of $\Lambda$ to the $G$-components of the holonomies and $(\Lambda_G)_*$ 
for the push-forward $F\rightarrow F\circ\Lambda_G^\inv$ of functions 
$F\in\cif(G^\ntg)$.

In view of Theorem \ref{symp} it is natural to ask if we can lift the mapping
class group action $\lambda\in\mapcld\mapsto
\Lambda\in\text{Diff}((\prgr)^\ntg)$ to an action 
$\lambda\in\mapcld\mapsto \tilde\Lambda\in\text{Diff}((\prgr)^\ntg)$
such that the following diagram commutes
\begin{equation}
\label{b1}
\begin{CD}
(\prgr)^{n+2g} @> \tilde\Lambda>> (\prgr)^{n+2g}\\
 @VV\tau V     @VV \tau V\\
(\prgr)^{n+2g}  @>\Lambda>>  (\prgr)^{n+2g}.\\
\end{CD}
\end{equation}

To define $\tilde\Lambda$ note that all generators of the mapping 
class group, defined by expressions \eqref{ad}-\eqref{etapb} 
and \eqref{braid} in the appendix, either leave the conjugacy 
classes $\calc_{\mu_is_i}$ associated to each puncture invariant 
or exchange the conjugacy classes of different punctures. Thus we 
can perform the following construction.

Let  $E$ be a group acting on $G^\ntg$ via 
$\xi\in E\mapsto \Xi_G\in\text{Diff}(G^\ntg)$, 
such that $E$ acts on the first $n$ copies of $G$ by 
conjugation and permutation
\begin{align}
\label{xiact}
\Xi_G: &u_\mi\mapsto \xi_\mi(u_{M_1},\ldots,u_{B_g})
\cdot u_{M_{\sigma(i)}} \cdot \xi_\mi^\inv(u_{M_1},\ldots,u_{B_g})\\
& u_{A_j}\mapsto \xi_{A_j}(u_{M_1},\ldots,u_{B_g}),\; u_{B_j}\mapsto 
\xi_{B_j}(u_{M_1},\ldots,u_{B_g}),\nonumber
\end{align}
with maps $\xi_{M_i},\xi_{A_j},\xi_{B_j}: G^\ntg \rightarrow G$ and a
permutation $ \sigma\in S_n$. Then 
\begin{align}
\label{tildedef}
\tilde\Xi_G: &v_\mi\mapsto \xi_{M_i}\circ\beta(v_{M_1},
\ldots,u_{B_g})\cdot v_{M_\sigma(i)}\\
& u_{A_j}\mapsto \xi_{A_j}\circ\beta (v_{M_1},\ldots,u_{B_g}),\;
u_{B_j}\mapsto \xi_{B_j}\circ\beta(v_{M_1},\ldots,u_{B_g}) \nonumber
\end{align}
defines 
 a group action $\xi\in E\rightarrow \tilde\Xi_G\in\text{Diff}(G^\ntg)$ 
and the definition \eqref{betadef} of the map $\beta$ implies
\bea
\label{betaid}
\beta\circ\tilde \Xi_G=\Xi_G\circ\beta\qquad\forall\xi\in E.
\eea 
This allows us to lift the action $\Lambda_G\in\text{Diff}(G^\ntg)$
of elements $\lambda\in\mapcld$ to an action
$\tilde\Lambda_G\in\text{Diff}(G^\ntg)$ related to $\Lambda_G$ via 
\eqref{betaid}.
We can extend  $\tilde\Lambda_G$ to a diffeomorphism $\tilde\Lambda$ on
$(\prgr)^\ntg$ by taking its action on $(\gothg^*)^\ntg$ to be the 
one defined by $\Lambda$,
which yields a mapping class group action  $\lambda\in\mapcld\mapsto
\tilde\Lambda\in\text{Diff}((\prgr)^\ntg)$ satisfying \eqref{b1}.

We can then use this lift of the mapping class group action on the 
flower algebra to an action on $(\prgr)^\ntg$ with symplectic potential
 \eqref{thetaundec}
to prove that the mapping class group action on the flower algebra is a 
Poisson action:
\begin{theorem} (Poisson action of the mapping class group)
\label{poissmapact}

The symplectic potential $\Theta$ \eqref{thetaundec} and the flower algebra
Poisson structure \eqref{poiss} on $(\prgr)^\ntg$  are invariant under
the mapping class group actions $\tilde\Lambda, \Lambda$, respectively.
\end{theorem}

{\bf Proof:}
For the symplectic potential $\Theta$ in \eqref{thetaundec} the
invariance under $\tilde\Lambda$
  can be shown by direct calculation using the expressions
 \eqref{ad}-\eqref{braid} for the action of the generators of $\mapcld$ 
on the curves $m_i$, $a_j$, $b_j$,
  the parametrisation \eqref{gparam} and the lift
  \eqref{tildedef}. The invariance of the flower algebra Poisson
  structure under $\lambda$ then follows from
  Theorem \ref{symp} and the
  commutative diagram \eqref{b1}, as we have for all $\lambda\in\mapcld$
  \begin{align}
\tau^* \Lambda_* \omega_\gothf=\tilde{\Lambda}_*\tau^* \omega_\gothf=
\widetilde{\Lambda}_* d\Theta=d\Theta=\tau^* \omega_\gothf.
\end{align}
Hence $\Lambda_*\omega_\gothf=\omega_\gothf$  by injectivity of the pullback
 with the surjective map $\tau$ in \eqref{taudef}.
A proof of the invariance of the Poisson bracket $\{\,,\,\}$ by direct
 calculation is given in \cite{we} for the case of the
  (2+1)-dimensional Poincar\'e group and can easily be extended to
  the case of a general group $\prgr$.\hfill $\Box$

\subsection{Infinitesimal generators for the action of Dehn twists}

After investigating the mapping class group action on the flower 
algebra and the Poisson manifold $(\prgr)^\ntg$ with symplectic potential
 \eqref{thetaundec}, we will now demonstrate that the action
of Dehn twists  can be related to an {\em
  infinitesimally generated} Poisson action of the group $G$.

Let $\gamma$ be an embedded i.~e.~non self-intersecting curve on the 
surface $S_{g,n}\mindee$ 
and let the same letter stand for the element of the 
(pure)
mapping class group given by the Dehn twist around $\gamma$ as outlined 
in the appendix. Denote by $\Gamma \in\text{Diff}((\prgr)^\ntg)$ and 
$\Gamma_G\in\text{Diff}(G^\ntg)$ the actions of this Dehn twist on the 
groups $(\prgr)^\ntg$ and $G^\ntg$, respectively, and by
$\tilde\Gamma\in\text{Diff}((\prgr)^\ntg)$ and $\tilde\Gamma_G
\in\text{Diff}(G^\ntg)$ their lifts according to
\eqref{tildedef} and \eqref{b1}. Parametrising the holonomy of the 
curve $\gamma$ as
$\text{Hol}(\gamma)=(u_\gamma,-\Ad^*(u_\gamma^\inv)\bj^\gamma)$ and 
expressing it
as a product of the holonomies $M_i,A_j,B_j$, we can introduce 
coordinate maps
$j^\gamma_a$  
analogous to \eqref{jmap}
\begin{align}
\label{jeta} 
j^\gamma_a\in\cif((\prgr)^\ntg): ({M_1},\ldots,M_n,A_1,\ldots,{B_g})
\mapsto j^\gamma_a,\qquad a=1,\ldots,\text{dim}\, G.
\end{align}

From the brackets \eqref{poiss} it follows that the coordinate functions 
$j^X_a$
generate a $G$-action on $\cif(G^\ntg)$ for all generators of the
fundamental group. One would like to generalise this statement to any
embedded curve $\gamma$. This requires one to find $G$-actions
 $\rho_\gamma: G\rightarrow\text{Diff}(G^\ntg)$,
 $\tilde\rho_\gamma: G\rightarrow
 \text{Diff}(G^\ntg)$ on $G^\ntg$ that are infinitesimally generated
 by $j^\gamma_a$ via these Poisson brackets
\begin{align}
\label{identdef2}
\{j_a^\gamma,F\}&=-\frac{\text{d}}{\text{dt}}|_{t=0}\;
F\circ\rho_\gamma(e^{-tJ^a})\qquad\forall
F\in\cif(G^\ntg)\\
\label{identdef}
\{j_a^\gamma,F\}_\Theta&=-\frac{\text{d}}{\text{dt}}|_{t=0}\;
F\circ\tilde\rho_\gamma(e^{-tJ^a})\qquad\forall
F\in\cif(G^\ntg).
\end{align}
Also, these $G$-actions should act on
 the group elements associated to the punctures by conjugation and
 left-multiplication, respectively, such that for each $g\in G$
 $\rho_\gamma(g),\tilde\rho_\gamma(g)\in\text{Diff}(G^\ntg)$ are
 related by \eqref{tildedef}. If such $G$-actions exist, they are uniquely
 defined by \eqref{identdef2}, \eqref{identdef}, since
 every element of the group $G$ can be written as a product of
 elements in the image of the exponential map.
Remarkably, it is possible to define such  $G$-actions $\rho_\gamma,
\tilde\rho_\gamma$
for each embedded curve $\gamma$ on $S_{g,n}\mindee$, and to relate them to 
the actions $\Gamma_G$, $\tilde\Gamma_G$ of the Dehn twist around
$\gamma$.

\begin{theorem}(Action of the Dehn twists on $G^\ntg$)
\label{gendt}

For any embedded curve $\gamma$ on the surface
$S_{g,n}\mindee$,  Eqs.~\eqref{identdef2} and \eqref{identdef} define associated 
$G$-actions
$\rho_\gamma,\tilde\rho_\gamma: G\rightarrow \text{Diff}(G^\ntg)$ 
related by \eqref{tildedef}, which conjugate and, respectively, 
left-multiply the group elements associated to the
punctures. In terms of these group actions,  the actions $\Gamma_G, \tilde\Gamma_G\in\text{Diff}(G^\ntg)$ of the
Dehn twist around $\gamma$ on $G^\ntg$ can be expressed as
\bea
\label{maprho}
\tilde \Gamma_G=\tilde \rho_\gamma(P_\gamma^\inv\circ\beta)\qquad \Gamma_G=\rho_\gamma(P_\gamma^\inv),
\eea
where
$P_\gamma^{\pm 1}: G^\ntg\rightarrow G$,
$P_\gamma^{\pm 1}:(u_{M_1},\ldots,u_{B_g})\mapsto u_{\gamma}^{\pm 1}$ maps to the (inverse of) the $G$-component of $\text{Hol}(\gamma)=(u_\gamma,-\Ad^*(u_\gamma^\inv)\bj^\gamma)$ expressed as a product in the $G$-components $u_{M_1},\ldots,u_{M_n},u_{A_1},\ldots, u_{B_g}$.
\end{theorem}

{\bf Proof: }

Because of identity \eqref{betaid} and the commutative diagram
\eqref{b1}, it is sufficient to prove the existence of such a $G$-action for the modified Poisson bracket $\{\,,\,\}_\Theta$ and
the action $\tilde\rho_\gamma$.

1.  As a first step, we show how a $G$-action $\tilde\rho_\gamma$
    associated to an embedded curve $\gamma$ that satisfies
    \eqref{identdef} and \eqref{maprho} gives rise to a corresponding
    group action $\tilde \rho_\xi$ for all curves $\xi$ that
    can be obtained from $\gamma$ via the action of $\mapcld$.
 Let $\lambda\in\mapcld$ be an
  element of the mapping class group with actions
  $\Lambda_G,\tilde\Lambda_G \in\text{Diff}(G^\ntg)$ on
  $G^\ntg$. Consider the (embedded) curve $\xi$ obtained from $\gamma$
    by acting with $\lambda$, write its
  holonomy as $\text{Hol}(\xi)=(u_\xi,
  -\Ad^*(u_\xi^\inv)\bj^\xi)$ and denote by
  $\Xi_G,\tilde\Xi_G\in\text{Diff}(G^\ntg)$ the actions of the Dehn
  twist around $\xi$ on $G^\ntg$. From the geometric definition of
    Dehn twists in the appendix it follows that the Dehn twists around
    the curves $\xi$ and $\gamma$ are related by
 $\xi=\lambda\circ\gamma\circ\lambda^\inv$. Using \eqref{grmapdef} and  the
definition  \eqref{tildedef} of the lifts, we deduce that the  associated actions $\tilde\Gamma_G$, $\tilde\Xi_G$ on $G^\ntg$
satisfy
$\tilde\Xi_G=\tilde\Lambda_G^\inv\circ\tilde\Gamma_G\circ\tilde\Lambda_G$.
On the other hand, the
  invariance of the symplectic potential $\Theta$ \eqref{thetaundec}
  under $\tilde\Lambda$
 implies
\bea
\label{thetainvar}
\{j^\xi_a,
F\}_\Theta=\{j^\gamma_a,F\circ\tilde\Lambda_G^\inv\}_\Theta\circ\tilde\Lambda_G\qquad\forall F\in\cif(G^\ntg),
\eea
since $j_a^\xi=_aj^\gamma\circ\tilde\Lambda$, and this allows us to define a $G$-action $\tilde\rho_\xi:
G\rightarrow\text{Diff}(G^\ntg)$ via
\begin{align}
\label{groupactgen}
&\tilde\rho_\xi(g)
:=\tilde\Lambda_G^\inv\circ\tilde\rho_\gamma(g)\circ \tilde\Lambda_G\qquad\forall g\in G.
\end{align}
Using the invariance of the Poisson bracket under the mapping class group and the
corresponding identity for $\gamma$, we see immediately that $\tilde\rho_\xi$
 satisfies \eqref{identdef}. Furthermore, since
 $\tilde\Lambda_G$ acts on the group elements associated to the
 punctures by left-multiplication and permutation, we see that
 $\tilde\rho_\xi$ acts on these elements by left-multiplication if the
 same is true for  $\tilde\rho_\gamma$.  
To prove that $\tilde\rho_\xi$ satisfies \eqref{maprho}, we note that the map $P^\inv_\xi$ is given by $P^\inv_\xi=P^\inv_\gamma\circ\Lambda_G$ and calculate
\begin{align}
\tilde\rho_\xi\left(P_\xi^\inv\circ\beta\right) (v_{M_1},\ldots u_{B_g}
)&=\tilde\rho_\xi\left(P_\xi^\inv\circ\beta (v_{M_1},\ldots u_{B_g}
)\right) (v_{M_1},\ldots u_{B_g}
)\\
&=
\tilde\rho_\xi(P_\gamma^\inv\circ\Lambda_G\circ \beta (v_{M_1},\ldots u_{B_g}
))
(v_{M_1},\ldots u_{B_g})\nonumber\\
&=\tilde\Lambda_G^\inv\circ\tilde\rho_\gamma(P_\gamma^\inv\circ\beta)\circ\tilde \Lambda_G(
v_{M_1},\ldots u_{B_g})\nonumber\\
&=\tilde\Lambda_G^\inv\circ\tilde\Gamma_G\circ\tilde\Lambda_G(
v_{M_1},\ldots u_{B_g})\nonumber\\
&=\tilde\Xi_G(v_\me,\ldots,u_\bff).\nonumber
\end{align}

2. We therefore only need to prove \eqref{maprho} for a set of
  curves containing one representative for  each orbit of the
  $\mapcld$-action on $\pi_1(S_{g,n}\mindee)$. Such a set of curves
  can be constructed using results from geometric
  topology \cite{geomtop}. It has been shown, see for example Lemma
  2.3.A. in \cite{geomtop}, that the equivalence classes of all
  non-separating curves $\gamma$ on the surface $S_{g,n}\mindee$, 
i.e. curves $\gamma$
  such that $(S_{g,n}\mindee)\setminus\gamma$ 
is connected, are in the same orbit
  under the action of the mapping class group. This is a consequence of the classification of
  two-dimensional surfaces via the Euler characteristic. We can apply the same
  argument to separating curves if we keep in mind that, unlike the handles, the punctures of our surface $S_{g,n}\mindee$
  can be distinguished via the conjugacy classes assigned to them. This allows us to conclude that any two separating curves $\gamma, \gamma'$ on
  $S_{g,n}\mindee $ such that the 
two components of $(S_{g,n}\mindee)\setminus\gamma$ and
  $(S_{g,n}\mindee)
\setminus\gamma'$
 contain the same number of handles and the same sets of punctures
lie in the same orbit under the action of the mapping class
  group. It is therefore sufficient to prove \eqref{maprho} for one
  non-separating curve, for example any of the
   curves in \eqref{twistcurves} except $\kappa_{\nu,\mu}$, and the
  separating curves $\gamma^{i_1\ldots i_r j_1\ldots j_s}$ pictured in Fig.~2.
\begin{align}
\label{sepcurves}
&\gamma^{i_1\ldots i_r j_1\ldots j_s}=[b_{j_s},a_{j_s}^\inv]\cdot[b_{j_{s-1}},
a_{j_{s-1}}^\inv]\cdots[b_{j_1},a_{j_1}^\inv]\cdot m_{i_r}\cdots
m_{i_1}\\
&\qquad 1\leq j_1<j_2<\ldots<j_s\leq j_{s+1}:=g,\;
1\leq i_1<i_2<\ldots<i_r\leq i_{r+1}:=n. \nonumber
\end{align}

\vbox{
\vskip .3in
\input epsf
\epsfxsize=12truecm
\centerline{
\epsfbox{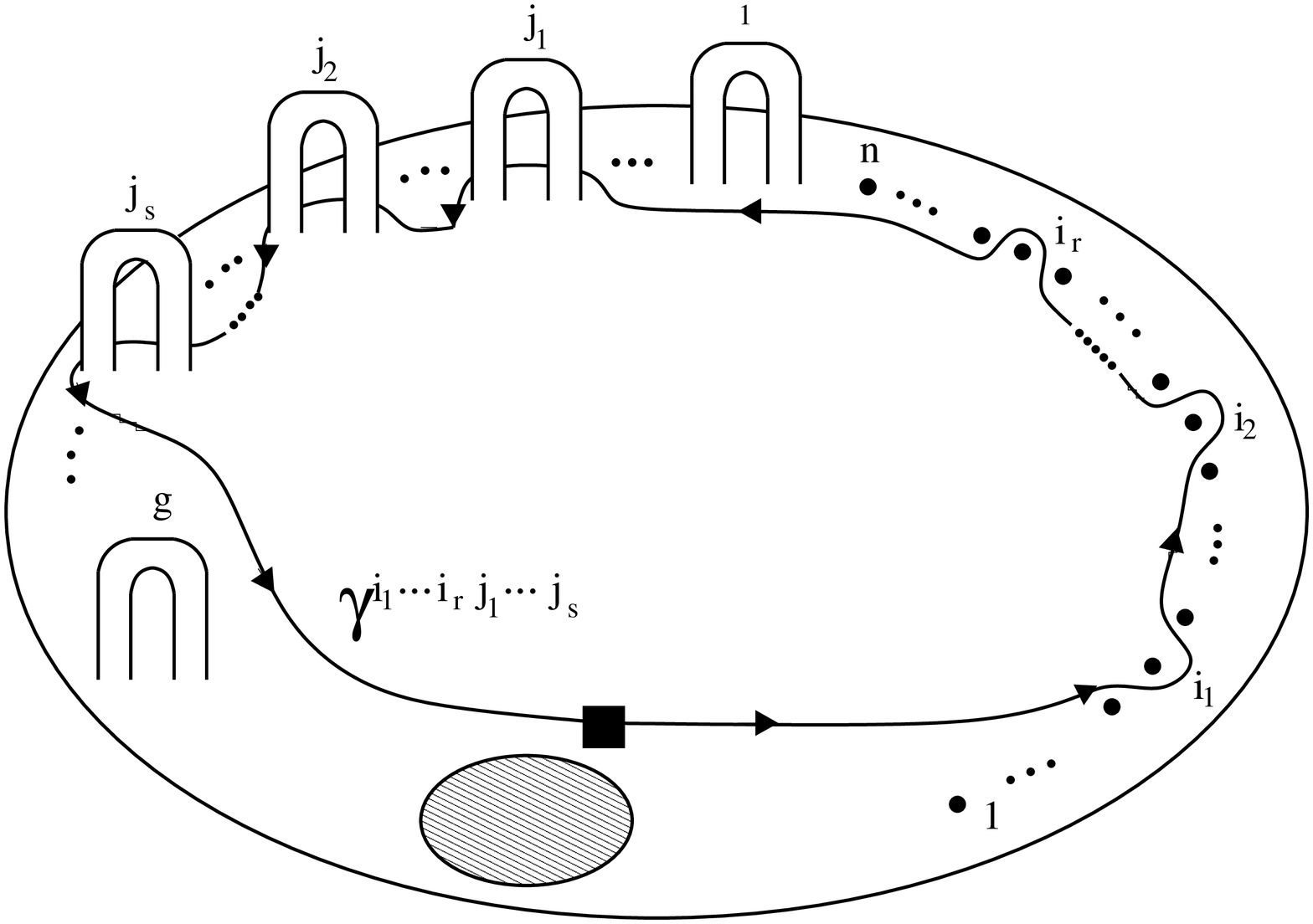}
}
\bigskip
{
\centerline{\bf Fig.~2 }
}
\centerline{\footnotesize Separating curves on the surface $S_{g,n}\mindee$}
\bigskip}

3. For a given curve $\gamma$ expressed as a product in
the generators $m_i, a_j,b_j \in\pi_1(S_{g,n}\mindee)$, we can parametrise
the holonomy as
$\text{Hol}(\gamma)=(u_\gamma,-\Ad^*(u_\gamma^\inv)\bj^\gamma)$
 and express it  in terms of
the holonomies $M_i$, $A_j$, $B_j$,  which allows us to 
calculate the Poisson brackets $\{j^\gamma_a, F\}_\Theta$
via \eqref{poiss}, \eqref{poisstheta}. For the set of
 curves \eqref{twistcurves} in the appendix, the holonomies are given
 by \eqref{ucurvedef}, \eqref{jcurvedef}, and we define the
 corresponding $G$-actions as 
\begin{align}
\label{gad}
\tilde\rho_{a_i}(g): &u_{A_i} \mapsto g u_{A_i} g^\inv\\
&u_\bi\mapsto [g,u_\ai]u_\bi g^\inv\nonumber\\
&u_X\mapsto [g,u_\ai] u_X [g,u_\ai]^\inv\quad\forall
X>\ai,\bi\nonumber\\
\label{gdd}
\tilde\rho_{\delta_i}(g): &u_\ai \mapsto [g,u_{\delta_i}]u_\ai g^\inv\\
&u_\bi \mapsto [g,u_{\delta_i}]u_\bi [g,u_{\delta_i}]^\inv\nonumber\\
&u_X \mapsto [g,u_{\delta_i}] u_X [g,u_{\delta_i}]^\inv\quad\forall
X>\ai,\bi\nonumber\\
\label{galpha}
\tilde\rho_{\alpha_i}(g): &u_\ai\mapsto [g,u_{\alpha_i}] u_\ai g^\inv\\
 &u_\bi\mapsto [g,u_{\alpha_i}] u_\bi [g,u_{\alpha_i}]^\inv\nonumber\\
 &u_{A_{i-1}}\mapsto g u_{A_{i-1}}\nonumber\\
 &u_{B_{i-1}}\mapsto g u_{B_{i-1}}g^\inv\nonumber\\
 &u_X\mapsto [g,u_{\alpha_i}] u_X
 [g,u_{\alpha_i}]^\inv\quad\forall X>\ai,\bi\nonumber\\
\label{geps}
\tilde\rho_{\epsilon_i}(g): &u_X\mapsto g u_Xg^\inv\qquad\forall X\in\{A_1,\ldots,B_{i-1}\}\\
 &u_\ai\mapsto [g,u_{\epsilon_i}] u_\ai g^\inv\nonumber\\
 &u_\bi\mapsto [g,u_{\epsilon_i}] u_\ai [g,u_{\epsilon_i}]^\inv\nonumber\\
 &u_X\mapsto [g,u_{\epsilon_i}] u_X [g,u_{\epsilon_i}]^\inv\qquad\forall X\in\{A_{i+1},\ldots,B_g\}\nonumber\\
\label{getapp}
\tilde\rho_{\kappa_{\nu,\mu}}(g): &v_{M_\nu}\mapsto g v_{M_\nu}\\
 &v_{M_\tau}\mapsto [g,u_{M_\nu}] v_{M_\tau}\qquad\forall\nu<\tau<\mu\nonumber\\
 &v_{M_\mu}\mapsto g v_{M_\mu}\nonumber\\
 &v_{M_\tau}\mapsto [g,u_{\kappa_{\nu,\mu}}] v_{M_\tau}\qquad\forall\tau>\mu\nonumber\\
&u_X\mapsto  [g,u_{\kappa_{\nu,\mu}}]u_X [g,u_{\kappa_{\nu,\mu}}]^\inv\qquad\forall X\in\{A_1,\ldots,B_g\}
\nonumber\\
\label{getapdelta}
\tilde\rho_{\kappa_{\nu, n+2i-1}}(g): &v_{M_\nu}\mapsto
g v_{M_\nu}\\
 &v_{M_\tau}\mapsto
[g, u_{M_\nu}] v_{M_\tau}\qquad\forall \tau=\nu+1,\ldots,n,\nonumber\\
 &u_{X}\mapsto
[g, u_{M_\nu}] u_X[g, u_{M_\nu}]^\inv\qquad\forall X\in\{A_1,\ldots,B_{i-1}\}\nonumber\\
 &u_\ai\mapsto
[g,u_{\kappa_{\nu, n+2i-1}}] u_\ai g^\inv\nonumber\\
 &u_\bi\mapsto
[g,u_{\kappa_{\nu, n+2i-1}}] u_\bi [g,u_{\kappa_{\nu, n+2i-1}}]^\inv\nonumber\\
 &u_{X}\mapsto
[g, u_{\kappa_{\nu,n+2i-1}}] u_X[g,
u_{\kappa_{\nu,n+2i-1}}]^\inv\;\forall
X\in\{A_{i+1},\ldots,B_{g}\}\nonumber\\
\label{getapb}
\tilde\rho_{\kappa_{\nu,n+2i}}(g): &v_{M_\nu}\mapsto g v_{M_\nu}\\
 &v_{M_\tau}\mapsto [g, u_{M_\nu}]v_{M_\tau}\qquad\forall\tau=\nu+1,\ldots,n \nonumber\\
 &u_X\mapsto [g, u_{M_\nu}]u_X[g, u_{M_\nu}]^\inv\qquad\forall X\in\{A_1,\ldots,B_{i-1}\}\nonumber\\
 &u_\ai\mapsto gu_\ai [g, u_{M_\nu}]^\inv\nonumber\\
 &u_\bi\mapsto gu_\bi g^\inv  \nonumber\\
 &u_X\mapsto [g, u_{\kappa_{\nu,n+2i}}]u_X[g, u_{\kappa{\nu,n+2i}}]^\inv\qquad\forall X\in\{A_{i+1},\ldots,B_{g}\}\nonumber,
\end{align}
where  $[\,,\,]$ denotes the group commutator on $G$ as given after \eqref{thetaundec},
$M_1<\ldots<M_n<A_1,B_1<\ldots<A_g,B_g$,
and $(u_{M_1},\ldots,u_{M_n},u_{A_1},\ldots,
u_{B_g})=\beta(v_{M_1},\ldots,v_{M_n},u_{A_1},\ldots,
u_{B_g})$. We listed only those
elements $u_X$ that transform non-trivially. It can be shown by direct
computation that expressions \eqref{gad}-\eqref{getapb} define
 $G$-actions on $G^\ntg$ which satisfy \eqref{identdef} and act on
the group elements associated to the punctures by left-multiplication. Furthermore,
comparing these $G$-actions with the action $\tilde\Gamma_G$ derived
  from expressions \eqref{ad}-\eqref{etapb} in the appendix, we see that they agree if we set $g=u_\gamma^\inv$.

Similarly, we calculate for the separating curves $\gamma^{i_1\ldots i_rj_1\ldots j_s}$ in
\eqref{sepcurves}
\begin{align}
\bj^{\gamma^{i_1\ldots i_r j_1\ldots
    j_s}}=&\bj^{M_{i_1}}+\Ad^*(u_{M_{i_1}})
\bj^{M_{i_2}}+\ldots+\Ad^*(u_{M_{i_r}}\cdots u_{M_{i_1}})\bj^{H_{j_1}}+\ldots \\
&+\Ad^*(u_{K_{j_{s-1}}}\cdots u_{K_{j_1}}u_{M_{i_r}}\cdots
    u_{M_{i_1}})\bj^{H_{j_s}}
\nonumber
\end{align}
with $u_{K_j}=[u_{B_j},u_{A_j}^\inv]=u_\bjj u_\aj^\inv u_\bjj^\inv u_\aj$ and
\bea
\label{jothi}\bj^{H_j}=(1-\Ad^*(u_\aj^{-1}u_\bjj^\inv u_\aj))\bj^\aj+(\Ad^*(u_\aj^{-1}u_\bjj^\inv
u_\aj)-\Ad^*(u_\bjj^{-1} u_\aj)\big)\bj^\bjj,
\eea
and define for all $g\in G$
\begin{align}
\label{sepgact}
\tilde\rho_{\gamma^{i_1\ldots i_rj_1\ldots j_s}}(g):\; &v_{M_{i_l}}\mapsto g
v_{M_{i_l}}\qquad  l=1,\ldots,r\\
 &u_{A_{j_l}}\mapsto g
u_{A_{j_l}} g^\inv,u_{B_{j_l}}\mapsto g
u_{B_{j_l}} g^\inv\qquad l=1,\ldots,s \nonumber\\
&v_\mi\mapsto [g, u_{M_{i_l}}\cdots u_{M_{i_1}}]v_\mi\qquad
\forall i_l<i<i_{l+1},\;l=1,\ldots,r\nonumber\\
&u_{A_j}\mapsto [g,u_{M_{i_r}}\cdots u_{M_{i_1}}]u_{A_j}
[g,u_{M_{i_r}}\cdots u_{M_{i_1}}]^\inv\qquad\forall 1\leq j<j_{1}\nonumber\\
&u_{B_j}\mapsto [g,u_{M_{i_r}}\cdots u_{M_{i_1}}]u_{B_j}
[g,u_{M_{i_r}}\cdots u_{M_{i_1}}]^\inv\qquad\forall 1 \leq j<j_{1}\nonumber\\
&u_{A_j}\mapsto [g, u_{K_{j_l}}\cdots u_{K_{j_1}}u_{M_{i_r}}\cdots u_{M_{i_1}}]u_{A_j} [g, u_{K_{j_l}}\cdots u_{K_{j_1}}u_{M_{i_r}}\cdots u_{M_{i_1}}]^\inv\nonumber\\
&\qquad\forall j_l<j<j_{l+1},\;l=1,\ldots,s\nonumber\\
&u_{B_j}\mapsto  [g, u_{K_{j_l}}\cdots
u_{K_{j_1}}u_{M_{i_r}}\cdots u_{M_{i_1}}]u_{B_j} [g, u_{K_{j_l}}\cdots
u_{K_{j_1}}u_{M_{i_r}}\cdots u_{M_{i_1}}]^\inv\nonumber\\
&\qquad
\forall j_l<j<j_{l+1},\;l=1,\ldots,s,\nonumber
\end{align}
Again, a straightforward calculation proves that \eqref{sepgact}
defines a $G$-action on $G^\ntg$ which satisfies \eqref{identdef} and
left-multiplies the group elements associated to the punctures.
After determining the action of the Dehn-twists around   
$\gamma^{i_1\ldots i_rj_1\ldots j_s}$ on the generators
of the fundamental group as described in the appendix, 
we can derive the associated actions $\tilde\Gamma^{i_1\ldots i_rj_1\ldots
  j_s}_G\in\text{Diff}(G^\ntg)$
\begin{align}
\label{sepdt}
 \tilde\Gamma^{i_1\ldots i_rj_1\ldots
  j_s}_G: \;
 &v_{M_{i_l}}\mapsto u_{\gamma^{i_1\ldots i_rj_1\ldots
  j_s}}^\inv
\cdot v_{M_{i_l}}\qquad  l=1,\ldots,r\\
 &u_{A_{j_l}}\mapsto  u_{\gamma^{i_1\ldots i_rj_1\ldots
  j_s}}^\inv\cdot
u_{A_{j_l}}\cdot u_{\gamma^{i_1\ldots i_rj_1\ldots
  j_s}}^\inv \qquad l=1,\ldots,s \nonumber\\
&u_{B_{j_l}}\mapsto u_{\gamma^{i_1\ldots i_rj_1\ldots
  j_s}}^\inv\cdot
u_{B_{j_l}}\cdot u_{\gamma^{i_1\ldots i_rj_1\ldots
  j_s}}\qquad l=1,\ldots,s \nonumber\\
&v_\mi\mapsto [ u_{\gamma^{i_1\ldots i_rj_1\ldots
  j_s}}^\inv, u_{M_{i_l}}\cdots u_{M_{i_1}}]\cdot v_\mi\qquad
\forall i_l<i<i_{l+1},\;l=1,\ldots,r\nonumber\\
&u_{A_j}\mapsto [ u_{\gamma^{i_1\ldots i_rj_1\ldots
  j_s}}^\inv, u_{M_{i_r}}\cdots u_{M_{i_1}}]\cdot u_{A_j}\cdot
 [ u_{\gamma^{i_1\ldots i_rj_1\ldots
  j_s}}^\inv, u_{M_{i_r}}\cdots u_{M_{i_1}}]^\inv\nonumber\\
&\qquad \quad\forall 1\leq j<j_{1}\nonumber\\
&u_{B_j}\mapsto [ u_{\gamma^{i_1\ldots i_rj_1\ldots
  j_s}}^\inv, u_{M_{i_r}}\cdots u_{M_{i_1}}]\cdot u_{B_j}\cdot 
[ u_{\gamma^{i_1\ldots i_rj_1\ldots
  j_s}}^\inv, u_{M_{i_r}}\cdots u_{M_{i_1}}]^\inv\nonumber\\
&\qquad\quad
\forall 1\leq j<j_{1}\nonumber\\
\intertext{}
&u_{A_j}\mapsto [ u_{\gamma^{i_1\ldots i_rj_1\ldots
  j_s}}^\inv, u_{K_{j_l}}\cdots u_{M_{i_1}}]\cdot u_{A_j}\cdot
 [ u_{\gamma^{i_1\ldots i_rj_1\ldots
  j_s}}^\inv, u_{K_{j_l}}\cdots u_{M_{i_1}}]^\inv\nonumber\\
&\qquad \quad\forall j_l<j<j_{l+1},\;l=1,\ldots,s\nonumber\\
&u_{B_j}\mapsto [ u_{\gamma^{i_1\ldots i_rj_1\ldots
  j_s}}^\inv, u_{K_{j_l}}\cdots u_{M_{i_1}}]\cdot u_{B_j}\cdot 
[ u_{\gamma^{i_1\ldots i_rj_1\ldots
  j_s}}^\inv, u_{K_{j_l}}\cdots u_{M_{i_1}}]^\inv\nonumber\\
&\qquad\quad
\forall j_l<j<j_{l+1},\;l=1,\ldots,s.\nonumber
\end{align}

We then see that they agree with \eqref{sepgact} if we set 
$g=u_{\gamma^{i_1\ldots i_rj_1\ldots
  j_s}}^\inv$. Therefore, group actions $\tilde\rho_\gamma:
  G\rightarrow \text{Diff}(G^\ntg)$ satisfying \eqref{identdef} and
  acting on the group elements associated to the punctures by
  left-multiplication exist for all embedded curves $\gamma$ on
  $S_{g,n}\mindee$ and 
are related to the action of the associated  Dehn twist on $G^\ntg$
  via \eqref{maprho},  which
   was to be shown.\hfill $\Box$

Theorem \ref{gendt} shows how the actions
$\Gamma_G,\tilde\Gamma_G\in \text{Diff}(G^\ntg)$ arising from the action
of a Dehn twists on the holonomies $M_i$, $A_j$, $B_j$
can be related to an action of the group $G$ that is infinitesimally
generated by the element $j^\gamma_a$ via the Poisson brackets
$\{\,,\,\}$ and $\{\,,\,\}_\Theta$.
This raises the question if the same can be said for the corresponding actions
$\Gamma, \tilde\Gamma\in\text{Diff}((\prgr)^\ntg)$. We show that this is the case by using the following lemma.
\begin{lemma}
\label{pushfwdlem}

Let $M$ be a manifold with diffeomorphism group $\text{Diff}(M)$ and denote by $\text{Vec}(M)$ the space of real vector fields on $M$. Let $\mathcal{L}$ be the infinite-dimensional Lie algebra $\mathcal{L}=\text{Vec}(M)\ltimes\cif(M)$ with Lie bracket
\begin{align}
\label{mliebr}
&[X,Y]_\mathcal{L}=[X,Y]_{Vec} & &[X,F]_\mathcal{L}=X.F & &[F,G]_\mathcal{L}=0 
\end{align} 
for $X,Y\in\text{Vec}(M)$, $F,G\in\cif(M)$, where
$X.F=\frac{\text{d}}{\text{dt}}|_{t=0}F(h^X_t(m))$ denotes the action of the vector
field $X$ on a function $F\in\cif(M)$ with flow $h_t^X$ generated
by $X$. Consider the action of diffeomorphisms $\phi\in\text{Diff}(M)$ on functions $F\in\cif(M)$ and vector fields $X\in\text{Vec}(M)$ via push-forward
\begin{align}
\label{pushfwd}
&\phi_*F=F\circ \phi^\inv & &(\phi_*X)F=\frac{\text{d}}{\text{dt}}|_{t=0}F(\phi\circ h^X_t\circ\phi^\inv).
\end{align}
Then, the push-forward with $\phi$ is a Lie algebra automorphism of $\call$ and
any Lie algebra isomorphism $\varphi:\call\mapsto \call$ with  $\varphi(\text{Vect}(M))\subset \text{Vect}(M)$ and $\varphi|_{\cif(M)}=\phi_*|_{\cif(M)}$ for some $\phi\in\text{Diff}(M)$ is equal to the push-forward with $\phi$: $\varphi=\phi_*.$
\end{lemma}

{\bf Proof:}
The first claim states simply the standard properties of the push-forward,
$\phi_*(X.F)=(\phi_*X).(\phi_*F)$ and $\phi_*[X,Y]_{Vec}=[\phi_*X,\phi_*Y]_{Vec}$
for $ X,Y\in\text{Vec}(M), F\in\cif(M)$, see \cite{AbMars, MarsRat}. The second follows from the fact that a vector field is uniquely determined by its action on functions: $(\varphi X).(\phi_* F)=\varphi(X.F)=\phi_*(X.F)=(\phi_* X).(\phi_*F)$.
\hfill$\Box$

Recalling the definition of the flower algebra and the modified bracket $\{\,,\,\}_\Theta$, we note that the subspace
$\gothg^\ntg\otimes\cif(G^\ntg)\oplus \cif(G^\ntg)$ with bracket $\{\,,\,\}_\Theta$ can be viewed as  a Lie algebra of type $\call$ in
Lemma \ref{pushfwdlem}. The Poisson brackets of generators  $j^X_a$ and functions
$F\in\cif(G^\ntg)$ allow us to identify the former with a basis of the
space of vector fields $\text{Vec}(G^\ntg)$.
 From the first set of brackets in \eqref{poiss}
it is then clear that the commutator of two vector fields
agrees with the Poisson bracket of
the associated elements in $\gothg^\ntg\otimes\cif(G^\ntg)$.

For any embedded curve $\gamma$
the associated Dehn twist
acts on $G^\ntg$ via the diffeomorphisms
$\Gamma_G,\tilde\Gamma_G\in\text{Diff}(G^\ntg)$ that map
$\text{Vec}(G^\ntg)$ to itself and act on functions $F\in\cif(G^\ntg)$ via
 the push-forward. We can therefore apply Lemma \ref{pushfwdlem} to
 the mapping class group action on the Lie algebra $\text{Vec}(G^\ntg)\ltimes\cif(G^\ntg)$ with Lie bracket $\{\,,\,\}_\Theta$. As the flower algebra is multiplicatively generated by the coordinate functions $j^X_a$ and functions in $\cif(G^\ntg)$, this defines the mapping class group action on the flower algebra with bracket $\{\,,\,\}_\Theta$ uniquely, and we see
that it is given by the push-forward with
 $\tilde\rho_\gamma(P_\gamma^\inv\circ\beta)$.
 Since $\Gamma$ and $\tilde
 \Gamma$ are related by the commutative diagram \eqref{b1} and
$\rho_\gamma$, $\tilde\rho_\gamma$ by \eqref{betaid}, the mapping
class group action on the flower algebra with bracket $\{\,,\,\}$
is then given by push-forward with $\rho_\gamma(P_\gamma^\inv)$.
Recalling from Lemma \ref{pushfwdlem} that the push-forwards with $\rho_\gamma(g)$ and $\tilde\rho_\gamma(g)$
define a Poisson action of the group $G$ on the flower algebra with
bracket $\{\,,\,\}$ and $\{\,,\,\}_\Theta$, we obtain the following theorem.

\begin{theorem} (Action of the Dehn twists on the flower algebra)
\label{algdt}

\begin{enumerate}
\item For any embedded curve $\gamma$ on $S_{g,n}\mindee$ the push-forward with  $\rho_\gamma(g),\tilde\rho_\gamma(g)\in \text{Diff}( G^\ntg)$ defines a Poisson action of the group $G$ on the flower
  algebra with bracket  $\{\,,\,\}$ and $\{\,,\,\}_\Theta$, respectively.
\item The actions $\Gamma$, $\tilde\Gamma$ of the associated Dehn twist on the flower algebra with bracket  $\{\,,\,\}$ and $\{\,,\,\}_\Theta$ are given by the push-forward with  $\rho_\gamma(P_\gamma^\inv)$ and  $\tilde\rho_\gamma(P_\gamma^\inv\circ\beta)$.
\end{enumerate}
\end{theorem}
One might ask if it is possible to define Hamiltonians such that the actions $\Gamma$, $\tilde\Gamma$ on the flower algebra are realised as their flow for some value of the flow parameter.
In the case where the exponential map $\exp:\gothg\rightarrow G$ is
surjective and elements $u\in G$ can be parametrised as
$u=\exp(p^aJ_a)$ such Hamiltonians can be given explicitly. We can then
write the holonomy along the curve $\gamma$ as
$\text{Hol}(\gamma)=(u_\gamma,-\Ad^*(u_\gamma^\inv)\bj^\gamma)$ with
$u_\gamma=\exp(p^a_\gamma J_a)$ and introduce maps
\bea
\label{pmaps}
p^a_\gamma\in\cif(G^\ntg): (u_{M_1},\ldots,u_\bff)\mapsto p^a_\gamma.
\eea
Considering the algebra element
\bea
\label{ceta}
c_\gamma=p^a_\gamma j^\gamma_a\in \gothg^\ntg\otimes\cif(G^\ntg).
\eea
and the one-parameter group of transformations $\phi^\gamma(t), \tilde\phi^\gamma(t)$ of the flower algebra generated by $c_\gamma$ via the Poisson brackets $\{\,,\,\}$ and $\{\,,\,\}_\Theta$
\begin{align}
\label{flowdef}
&\frac{\text{d}}{\text{dt}}|_{t=0}\,\phi^\gamma(t)\chi=\{c_\gamma,\chi\} &
 &\frac{\text{d}}{\text{dt}}|_{t=0}\tilde\phi^\gamma(t)\chi=\{c_\gamma,\chi\}_\Theta\\
&\forall \chi\in S\left(\bigoplus_{k=1}^{n+2g}\gothg\right)\otimes\cif(G^{n+2g}),\nonumber
\end{align}
 we obtain
\bea
\{c_\gamma, F\}=-p^a_\gamma\frac{\text{d}}{\text{dt}}|_{t=0}\,
F\circ\rho_\gamma(e^{-tJ_a})=\frac{\text{d}}{\text{dt}}|_{t=0}\,
F\circ\rho_\gamma(e^{tp_\gamma^a J_a})\qquad\forall F\in\cif(G^\ntg)
\eea
and an analogous expression involving $\{\,,\,\}_\Theta$ and
$\tilde\rho_\gamma$. This implies
\begin{align}
\label{fident}
&\phi^\gamma(1) F=F\circ\Gamma_G^\inv & &\tilde\phi^\gamma(1) F=F\circ\tilde\Gamma_G^\inv.
\end{align}
Furthermore, it follows from
\bea
\{c_\gamma,j^X_a\}=p_\gamma^b\{j^\gamma_b, j^X_a\}+j^\gamma_b\{p_\gamma^b,j^X_a\}\qquad\forall X\in\{M_1,\ldots,M_n,A_1,\ldots,B_g\}
\eea
and the structure of the expression for $j^\gamma_b$ in terms of the coordinate functions $j^Y_b$, $Y\in\{M_1,\ldots,B_g\}$, associated to the generators of the fundamental group that $\{c_\gamma,j^X_a\}$ is a linear combination these coordinate functions $j^Y_b$ with coefficients in $\cif(G^\ntg)$. The identification of the coordinate functions $j^Y_b$ with vector fields on $G^\ntg$, discussed after Lemma \ref{pushfwdlem}, then implies that $\phi^\gamma$ maps $\text{Vec}(G^\ntg)$ to itself.
\begin{theorem} 
The one-parameter groups of transformations $\phi^\gamma(t)$, $\tilde\phi^\gamma(t)$ act as Poisson isomorphisms on the flower algebra with bracket $\{\,,\,\}$ and  $\{\,,\,\}_\Theta$, respectively, and agree with the action of the associated Dehn twist at $t=1$
\begin{align}
&\phi^\gamma(1)=\Gamma_* & &\tilde\phi^\gamma(1)=\tilde\Gamma_*.
\end{align}
\end{theorem}

{\bf Proof:}
That the action of  $\phi^\gamma(t)$, $\tilde\phi^\gamma(t)$ is a
Poisson action $\forall t\in\mathbb{R}$ follows from the fact that
they are infinitesimally generated via the Poisson brackets
$\{\,,\,\}$ and  $\{\,,\,\}_\Theta$ \cite{AbMars, MarsRat}. That they agree with the action of the Dehn twists $\gamma$, $\tilde\gamma$ at $t=1$ can be deduced from \eqref{fident}, the fact that they are Poisson actions and that they map the space $\text{Vec}(G^\ntg)$ to itself by using Lemma \ref{pushfwdlem}.\hfill $\Box$

\section{The quantum action of the mapping class group}

In this section we investigate the action of the mapping class group
 on the quantised flower algebra and its representation spaces defined
 in Def.~\ref{qflower}, Def.~\ref{flowrep}.
For the case of Chern-Simons theory with compact, semisimple gauge
 groups the corresponding quantum action of the mapping class group
 has been studied by Alekseev and Schomerus
 \cite{AS}, who claim that elements of the mapping class
 group act as algebra automorphism on the quantum algebra constructed
 via their formalism of combinatorial quantisation of Chern-Simons
 theory\footnote{ The result was announced in \cite{AS} but the proof does not appear to have been published.}. Furthermore, they relate this
 quantum action of the mapping class group to an action of a quantum group
 associated to the gauge group of the underlying Chern-Simons
 theory. In terms of this quantum group action, Dehn twists around
 embedded curves are given
 by the action of the ribbon element and the exchange of punctures 
  is implemented  via the universal $R$-matrix.

We generalise and prove these results for semidirect product gauge 
groups of type
$\prgr$. Using the fact that elements of the mapping class group act
as Poisson isomorphisms on the classical flower algebra and the rather
close relation between classical and quantised flower
algebra, we prove that the mapping class group acts on the quantised
flower algebra via algebra isomorphisms. We show that this mapping class group
action on the quantised flower algebra can be implemented as an action
on its representation spaces. Finally, we
relate the mapping class group action to representations of a quantum
group, which in our case is the quantum double $D(G)$ of the group $G$.

In the proof of Theorem 4.1. in  \cite{we2}, we demonstrated that any
 Poisson isomorphism of the flower algebra that maps the subspace
 $\gothg^\ntg\otimes\cif(G^\ntg)\oplus\cif(G^\ntg)$ to itself gives
 rise to an algebra isomorphism of the quantised flower algebra
 \eqref{flowernew}. As the quantised flower algebra is
 multiplicatively generated by elements of $\cif(G^\ntg)$ and
 $\gothg^\ntg\otimes\cif(G^\ntg)$, this algebra isomorphism is
 uniquely defined by its action on the subspace
 $\gothg^\ntg\otimes\cif(G^\ntg)\oplus\cif(G^\ntg)$. Furthermore, 
if we identify the
 isomorphic subspaces $\gothg^\ntg\otimes\cif(G^\ntg)\oplus\cif(G^\ntg)$
 in the classical and quantised flower algebra, the quantum action on
 this subspace agrees with the classical action. Both the generators
 \eqref{ad}-\eqref{etapb} of the pure mapping class group $\pmapcld$
 and the generators \eqref{braid}
 satisfy the condition above, so that the classical action of the
 mapping class group gives rise to a mapping class group action as
 algebra automorphisms of the quantised flower algebra. The results
 in Sect.~\ref{classmapact} then allow us to implement this action on 
the representation spaces in Theorem \ref{flowrep}. The key 
observation is that the states $\psi\in V_\repind$ in the Hilbert 
spaces defined in Def.~\ref{flowrep} are vector-valued functions 
on $G^\ntg$ satisfying an equivariance condition. Each component 
of $\psi$ may thus be viewed as an element of the flower algebra with 
bracket $\{\,,\,\}_\Theta$. This allows us to define a representation 
of $\mapcld$ on the Hilbert spaces $V_\repind$ by extending its action 
on the flower algebra componentwise to $\psi\in V_\repind$.

\begin{theorem} (Quantum action of the mapping class group)
\label{qmapact}

\begin{enumerate}
\item Elements $\lambda\in\mapcld$ of the mapping class group act as algebra automorphisms $\hat\Lambda: \hat\gothf\rightarrow \hat\gothf$ on the quantum flower algebra.
\item Let $\pi_\lambda\in S_n$ be the permutation associated to $\lambda$ via the map $\pi$ in \eqref{sequen} and
\bea
\label{tdef}
p_\lambda:\,V_{s_1}\otimes\ldots\otimes V_{s_n}\rightarrow V_{s_{\pi_\lambda(1)}}\otimes\ldots\otimes V_{s_{\pi_\lambda(n)}}
\eea
the map which permutes the factors in the tensor product. Then the map
\bea
\label{repmap}
L_\lambda:\;\psi\in V_\repind\;\mapsto\; p_\lambda\circ\psi\circ\tilde\Lambda_G^\inv \in V_{\mu_{\pi_\lambda(1)}s_{\pi_\lambda(1)}\ldots \mu_{\pi_\lambda(n)}s_{\pi_\lambda(n)}},
\eea
defines a representation of the mapping class group on the Hilbert
spaces $V_\repind$ given in Theorem \ref{flowrep}. On the
dense subspace $V^\infty_\repind$ carrying the
representations of the quantised flower algebra, it satisfies
\bea
\label{repident}
\Pi_\repind(\hat\Lambda\chi)= L_\lambda\circ\Pi_\repind(\chi)\circ L_\lambda^\inv\qquad\forall\chi\in\hat\gothf.
\eea
\end{enumerate}
\end{theorem}

{\bf Proof:} The first claim follows from the discussion in \cite{we2} as explained above. 
 That \eqref{repmap} defines a representation of the mapping class
 group is a consequence of the properties of the push-forward. To prove identity \eqref{repident}, we use the fact that the action of the mapping class group on the quantum flower algebra is uniquely defined by its action on functions $F\in\cif(G^\ntg)$ and the generators $j^X_a$, on which it agrees with the corresponding classical action. With expression \eqref{quantact} for the action of these elements on the representation spaces $V_\repind$, we obtain
\begin{align}
\label{prrepid1}
\Pi_\repind(\hat\Lambda (1\otimes
F))L_\lambda\psi\;\;&=\left(((\Lambda_G)_* F)\circ\beta\right)\cdot
L_\lambda\psi=((\tilde \Lambda_G)_* (F\circ\beta))\cdot L_\lambda\psi\nonumber\\
&=L_\lambda (\Pi_\repind (1\otimes F)\psi),
\end{align}
where we used the fact that $\Lambda_G$ and $\tilde \Lambda_G$ are related by equation \eqref{betaid}. Recalling the definition of the action $\tilde\Lambda$ on the classical flower algebra and the fact that this action is a Poisson action by Theorem \ref{poissmapact}, we calculate for the action of elements $j^X_a\otimes F\in\gothg^\ntg\otimes\cif(G^\ntg)$
\begin{align}
\label{prrepid2}
&\Pi_\repind(\hat\Lambda (j^X_a\otimes F)) L_\lambda\psi=-i\hbar ((\Lambda_G)_*F)\circ\beta \cdot \{\tilde\Lambda_* j^X_a, L_\lambda\psi\}_\Theta\\
&\qquad=-i\hbar (\tilde \Lambda_G)_*(F\circ\beta)\cdot L_\lambda\{j^X_a, \psi\}_\Theta= L_\lambda (\Pi_\repind(j^X_a\otimes F)\psi)\nonumber,
\end{align}
which together with \eqref{prrepid1} proves the claim.\hfill $\Box$

We would now like to relate this action of the mapping class group on the
Hilbert spaces $V_\repind$ to different representations of a quantum
group associated to the gauge group $\prgr$, generalising the corresponding
result for compact, semisimple gauge groups obtained by Alekseev and Schomerus \cite{AS}. Whereas the quantum group
representations and their relation to elements of the mapping class
group are stated rather implicitly there, we find that our
formulation admits an explicit constructing relating them to the
classical structures discussed in Sect.~\ref{classmapact}.

The quantum group relevant to our formulation is the quantum double
$D(G)$ of the group $G$.
Using the definition given in \cite{we2},
we can identify the quantum double as a vector space with the space of continuous
functions on $G\times G$ with compact support: $D(G) =C_0(G\times G,\CC)$. In order to exhibit the structure of $D(G)$ as a ribbon-Hopf-*-algebra, it is necessary to introduce Dirac delta functions which are not strictly in
$C_0(G\times G,\CC)$ but can be included by simply adjoining
them.
Thus we define   multiplication $\bullet$, identity 1,
co-multiplication $\Delta$,
co-unit $\epsilon$, antipode $S$ and involution ${}^*$ via
\bea
\label{algebra}
(F_1\bullet F_2)(v,u)&:=&\int_G F_1(z,u)\,F_2(z^{-1}v,z^{-1}uz)\,dz
 \\
1(v,u)&:=&\delta_e(v) \nonumber\\
(\Delta F)(v_1,u_1;v_2,u_2)&:=&F(v_1,u_1u_2)\,\delta_{v_1}(v_2)\nonumber \\
\epsilon(F)&:=&\int_G F(v,e)\,dv\nonumber \\
(S F)(v,u)&:=&F(v^{-1},v^{-1}u^{-1}v)\nonumber\\
F^*(v,u)&:=&\overline{F(v^{-1},v^{-1}uv)}.\nonumber
\eea
The universal
$R$-matrix is then given by
\bea
\label{univrdef}
R(v_1,u_1;v_2,u_2) = \delta_e(v_1)\delta_e(u_1v_2^{-1})
\eea
and the central ribbon element $c$ by
\bea
\label{randc}
c(v,u) =  \delta_v(u).
\eea

We start by considering Dehn twists around embedded curves
$\gamma$. In order to relate the
action of these Dehn twists to the action of the ribbon element in
 representations of $D(G)$, we need to find a way of associating such
representations of $D(G)$ to each of the curves $\gamma$. In view of the
classical results in Sect.~\ref{classmapact}, one could expect
these representations to involve the $G$-action
$\tilde\rho_\gamma$ and the map $P_\gamma:G^\ntg\rightarrow G$. To pursue this
intuition further, we note that, given an action of the group $G$ on a manifold $M$ together with map
$\phi: M\rightarrow G$ satisfying a certain compatibility condition, there is a canonical way of constructing representations of the
 quantum double $D(G)$ on the space $L^2(M)$:

\begin{lemma} \label{qugrrepm}
Let $G$ be a unimodular Lie group with a continuous action $\rho: g\in
G\rightarrow \text{Diff}(M)$ on a manifold $M$ equipped with a 
 Borel measure $dm$ invariant under the $G$-action $\rho$. Let $\phi: M\rightarrow
G$ be a continuous map satisfying the equivariance condition
\bea
\label{phirhoid}
\phi(\rho(g)m)=g\cdot \phi(m)\cdot g^\inv\qquad\forall g\in G, m\in M.
\eea
Then a unitary representation $\Pi_{\rho,\Phi}$ of the quantum double $D(G)$ on the space $L^2(M)$ is
given by
\bea
\label{repform}
\Pi_{\rho,\phi}(F)\psi\,(m)=\int_G\, F(z,\phi(m))
\psi\circ\rho(z^\inv)(m) dz,
\eea
where $dz$ is the Haar measure on $G$.
\end{lemma}

{\bf Proof:} That \eqref{repform} defines a unitary
representation of $D(G)$  can be shown by direct calculation using the
definition \eqref{algebra} of the quantum double $D(G)$ and the
compatibility condition \eqref{phirhoid}.
 \hfill$\Box$

In our situation, we have $M=G^\ntg$ and consider the $G$-action $\rho=\tilde\rho_\gamma$ and the map
$\phi=P_\gamma^\inv\circ\beta$. We need to show that they satisfy
  \eqref{phirhoid}. For the set of curves
 \eqref{twistcurves} in the appendix and the separating curves \eqref{sepcurves}, this follows directly from
 identity \eqref{betaid} and expressions \eqref{gad}-\eqref{getapb} and \eqref{sepgact} for the $G$-actions.  Identity
 \eqref{groupactgen} then allows us to generalise this result to all
 embedded curves on $S_{g,n}\mindee$ as follows. If   $\tilde\rho_\gamma$ and
$P_\gamma^\inv\circ\beta$ satisfy \eqref{phirhoid} for an embedded curve
$\gamma$ and  $\xi$ is obtained from
$\gamma$ via the action of an element $\lambda\in\mapcld$ of the mapping class group, we have
\begin{align}
P_\xi^\inv\circ\beta \circ\tilde\rho_\xi(g)
&=P_\gamma^\inv\circ\Lambda_G\circ\beta\circ\tilde\Lambda_G^\inv
\circ\tilde\rho_\gamma(g)\circ \tilde\Lambda_G\\
&=P_\gamma^\inv\circ\beta\circ \tilde\rho_\gamma(g)\circ\tilde\Lambda_G\nonumber\\
&=g\cdot (P_\gamma^\inv\circ\beta\circ\tilde\Lambda_G)\cdot g^\inv\nonumber\\
&=g\cdot( P_\gamma^\inv\circ\Lambda_G\circ\beta)\cdot g^\inv\nonumber\\
&=g\cdot (P_\xi^\inv\circ\beta)\cdot g^\inv,\nonumber
\end{align}
so that \eqref{phirhoid} holds for $\tilde\rho_\xi$ and
$P_\xi^\inv\circ\beta$ as well. The curves \eqref{twistcurves} and
\eqref{sepcurves} are such that, up to homotopy, every embedded curve on
$S_{g,n}\mindee$ can be obtained by acting on one of them with
$\mapcld$. Hence the result \eqref{phirhoid} holds for all embedded curves.

By Lemma \ref{qugrrepm}, the $G$-action $\tilde\rho_\gamma$ and the map
$P_\gamma^\inv\circ\beta$ then define a unitary representation of the quantum
double $D(G)$ on the Hilbert spaces $V_\repind$. Using expression
\eqref{randc} for the ribbon element, we see that its representation
on the space $V_\repind$ agrees with the action of the
corresponding Dehn twist:

\begin{theorem}(Quantum action of Dehn twists)

\begin{enumerate}
\item For any embedded curve $\gamma$ on $S_{g,n}\mindee $ the map 
\begin{align}
\label{dtrepdouble}
&\Pi_\gamma(F)\psi(v_{m_1},\ldots,v_{M_n},u_\aee,\ldots,u_\bff)=\int_G F(z,u_\gamma^\inv)\,\psi\circ\tilde\rho_\gamma(z^\inv)(v_{M_1},\ldots,u_\bff)dz\nonumber\\
&\qquad=\int_G F\left(z,P_\gamma^\inv\circ\beta(v_{M_1},\ldots,u_\bff)\right)\, \psi\circ\tilde\rho_\gamma(z^\inv)(v_{M_1},\ldots,u_\bff)dz
\end{align}
defines a unitary representation of the quantum double $D(G)$ on the Hilbert space $V_\repind$.
\item The action of  the central ribbon element $c\in D(G)$ \eqref{randc} on $V_\repind$ agrees with the mapping class group action defined in Theorem \ref{qmapact}
\bea
\label{dtdoubleid}
\Pi_\gamma(c)\psi=(\tilde\Gamma_G)_*\psi \qquad\forall\psi\in V_\repind.
\eea
\end{enumerate}
\end{theorem}

To find the representations associated to the generators \eqref{braid}
of the braid group, 
we use  the standard result that representations 
of a quantum group give rise to
representations of the braid group via the universal $R$-matrix
\cite{CP}. If we associate a representation of the quantum double
$D(G)$ to each puncture of the surface $S_{g,n}$ as in \cite{we2}, 
the universal
$R$-matrix of $D(G)$ acts on the tensor product of two such
representations. The following theorem generalises results of \cite{BM,KBM}.

\begin{theorem}(Quantum action of the braid group)

\label{braidhilbact}
Define representations $\Pi_{\mu_is_i}: D(G)\rightarrow
Hom(V_\repind,V_\repind)$  of $D(G)$ by
\begin{align}
\label{rmatrep}
&\Pi_{\mu_i s_i}(F)\psi(v_\me,\ldots,v_\mf, u_\aee,\ldots,u_\bff)\\
&\qquad\qquad=\int_G F(z, v_{M_i}g_{\mu_i}v_{M_i}^\inv)\psi(v_\me,\ldots,v_{M_{i-1}},z^\inv v_{M_i},v_{M_{i+1}},\ldots, u_\bff)dz,\nonumber
\end{align} and let $\pi^i: G^\ntg\rightarrow G^\ntg$ be the map that exchanges the $i^{th}$ and $(i+1)^{th}$ copy of $G$
\begin{align}
\label{flipdef}
\pi^{i}: (v_\me,\ldots,v_\mf,u_\aee,\ldots,u_\bff)\mapsto (v_\me,\ldots,v_{M_{i+1}},v_{M_{i}},\ldots    v_\mf,u_\aee,\ldots,u_\bff)
\end{align}
Then, the action of the generators \eqref{braid} of the braid group on  $V_\repind$ is given by
\begin{align}
\label{braidrid}
L_{\sigma^i}\psi=p^{i}((\Pi_{\mu_is_i}\otimes\Pi_{\mu_{i+1}s_{i+1}})(R)) \circ \pi^{i}_*\psi\qquad\forall \psi\in V_\repind, 
\end{align}
where $p^{i}: V_\repind\rightarrow V_{s_1}\otimes\ldots\otimes
V_{s_{i-1}}\otimes V_{s_{i+1}}\otimes V_{s_i}\otimes
V_{s_{i+2}}\otimes\ldots\otimes V_{s_n}$ exchanges the spaces $V_{s_i}$ and $V_{s_{i+1}}$ in the tensor product.
\end{theorem}

{\bf Proof:} That \eqref{rmatrep} defines a representation of the
quantum double $D(G)$ on $V_\repind$ can be verified by direct
calculation using \eqref{algebra}. To prove \eqref{braidrid}, we insert
the definition \eqref{univrdef} of the universal $R$-matrix in
\eqref{rmatrep} and obtain
\begin{align}
\label{proofcalc}
&p^{i}(\Pi_{\mu_i s_i}\otimes\Pi_{\mu_{i+1} s_{i+1}})(R) \left(
  \pi^{i}_*\Psi\right)(v_\me,\ldots, u_\bff)\\
&\qquad=p^{i}(\Pi_{\mu_i
  s_i}\otimes\Pi_{\mu_{i+1} s_{i+1}})(R)\psi(v_\me,\ldots,v_{M_{i+1}},v_\mi,\ldots,u_\bff)\nonumber\\
&\qquad=p^{i}\int_{G\times G} R(z_1, v_{M_{i+1}} g_{\mu_{i+1}} v_{M_{i+1}}^\inv,
  z_2, v_{\mi} g_{\mu_{i}} v_{\mi}^\inv )\psi(v_\me,\ldots,z_1^\inv v_{M_{i+1}},
  z_2^\inv v_\mi,\ldots,u_\bff)\nonumber\\
&\qquad=p^{i}\psi(v_\me,\ldots,v_{M_{i-1}}, v_{M_{i+1}}, (v_{M_{i+1}}g_{\mu_{i+1}} v_{M_{i+1}})^\inv\cdot v_\mi, v_{M_{i+2}},\ldots,u_\bff).\nonumber
\end{align}
Recalling the definition of $\tilde\sigma^i_G$ via \eqref{braid} and \eqref{tildedef}, we see that this agrees with $L_{\sigma^i}\psi$.\hfill$\Box$

\section{Concluding remarks}
In this paper we constructed a Poisson action of the mapping class
group $\mapcld$ on the flower algebra and on the representation 
spaces of the associated quantum algebra. We related the classical action
of Dehn twists to an infinitesimally
generated $G$-action and, in the case where the exponential map is 
surjective, to Hamiltonian flows of certain conjugation invariant
functions on  $\prgr$. In the quantum theory, we showed how the 
mapping class group representation can be expressed in terms of 
 the ribbon element and universal
$R$-matrix of the quantum double $D(G)$.
Our results were derived  for any connected, 
simply-connected  and unimodular
 finite-dimensional Lie group $G$,
but the assumptions of connectedness and unimodularity 
can be dropped at the expense of  mild technical complications.

We feel that the mathematical structure of the flower algebra makes it an
object of investigation in its own right. However, it attracted 
our attention because of  its relevance to physics, more precisely, 
its role in the description of the phase space of (2+1)-dimensional
gravity in the Chern-Simons formulation. 
The phase space of Chern-Simons theory with gauge group $\prgr$, 
the moduli space of flat $\prgr$-connections,
 can be obtained from our set of 
holonomies $M_1,\ldots,M_n,A_1,B_1,\ldots A_g,B_g$
by imposing  the condition \eqref{pirel} 
and dividing by the $\prgr$-action
which simultaneously  conjugates all holonomies.
Formal arguments, based on the analogy with the discussion
in \cite{AS}, suggest that the Poisson action of $\mapcld$
on the flower algebra  descends to a symplectic action 
of $\mapcl$ on the moduli space.
A mathematically rigorous implementation of these arguments
for  the non-compact
groups $\prgr$  considered here would be interesting, 
particularly 
for the physically relevant cases
where $\prgr$ is the
 universal cover of  the three-dimensional Euclidean or
Poincar\'e group.

In the quantum theory,  the 
classical conjugation symmetry is replaced by  an 
action of the quantum double $D(G)$ on the Hilbert  spaces $V_\repind$, 
see \cite{we}, in particular Eq.~(4.27).
The constraint \eqref{pirel} is then implemented on these Hilbert 
spaces by imposing invariance under the action of the quantum double.
The action of $\mapcld$  on $V_\repind$ 
derived  in Sect.~4 of the present paper  commutes
with this action of the quantum double. Formally, one 
obtains an action of the 
mapping class group $\mapcl$ 
on the invariant states in $V_\repind$, but 
there are again technical difficulties related to the 
non-compactness of $\prgr$: the states which are invariant under 
the $D(G)$-action are singular and not proper elements of $V_\repind$.
For the case of Chern-Simons theory with the non-compact but semisimple 
gauge group  $SL(2,\CC)$, a mathematically rigorous way of defining 
invariant states has been derived in \cite{BNR}.
It would be interesting to see if a similar method can be applied to our  
situation and  to investigate
the resulting mapping class group action on the reduced Hilbert space, 
again with particular attention to 
the  three-dimensional 
Euclidean  or Poincar\'e group.


\section*{Acknowledgements}
We thank Jim Howie for valuable information concerning the orbits of the mapping class group action on the fundamental groups of surfaces. CM acknowledges financial support by the Engineering and Physical Sciences Research Council and  a living stipend from the Studienstiftung des Deutschen Volkes. BJS acknowledges an Advanced Research Fellowship of the 
 Engineering and Physical Sciences Research Council.

\newpage

\appendix
\section{The generators of the mapping class group}

For the convenience of the reader we give a set of generators of the mapping
class groups $\mapcl$ and $\mapcld$
 with explicit formulae for their actions on a
set of generators of the fundamental groups
$\pi_1(S_{g,n})$ and $\pi_1(S_{g,n}\mindee)$. 
Some of the contents of
this appendix are taken from our discussion in \cite{we}.

The pure mapping class group
$\pmapcld$ of the punctured surface $S_{g,n}\mindee$ is the  
subgroup of $\mapcld$ which leaves each puncture
fixed. A set of generators and defining relations has been derived by
Birman \cite{birmgreen, birmorange} but for us the set of generators used
 by Schomerus and Alekseev \cite{AS} is more convenient, which was
first given in \cite{Wajnryb}, see also \cite{MatPol}. Note that
this set also generates $\pmapcl$, with additional relations.

The generating  set consists of
 Dehn twists around the curves $a_i$, $\delta_i$,
$\alpha_i$, $\epsilon_i$, $\kappa_{\nu,\mu}$, $\kappa_{\nu,n+2i-1}$
 and $\kappa_{\nu,n+2i}$ pictured in Fig.~3.

\vbox{
\vskip .3in
\input epsf
\epsfxsize=10truecm
\centerline{
\epsfbox{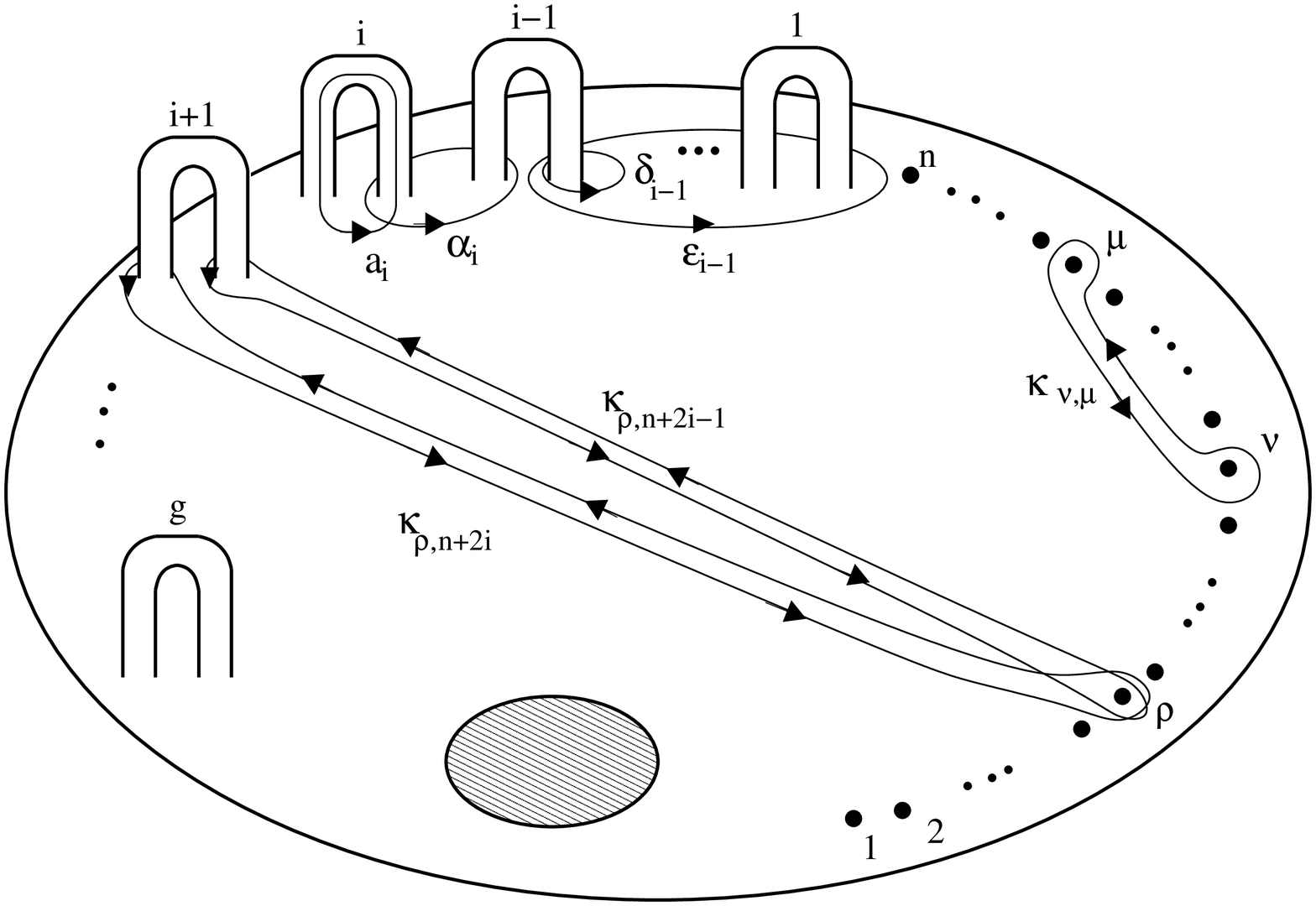}
}
\bigskip
{
\centerline{\bf Fig.~3 }
}
\centerline{\footnotesize The curves associated to the generators of the (pure)
  mapping class group $\pmapcld$}
\bigskip}

They can be expressed 
in terms of the generators $m_1,\ldots,m_n,a_1,b_1,\ldots,a_g,b_g$ of
the fundamental group shown in Fig.~1 as follows: 
\begin{align}
\label{twistcurves}
&a_i &  &i=1,\ldots,g\\
&\delta_i=a_i^{-1}b_i^{-1}a_i & &i=1,\ldots,g\nonumber\\
&\alpha_i=a_i^{-1}b_i^{-1}a_ib_{i-1} & &i=2,\ldots,g\nonumber\\
&\epsilon_i=a_i^{-1}b_i^{-1}a_i
\cdot(b_{i-1}a^{-1}_{i-1}b^{-1}_{i-1}a_{i-1})
\cdot\ldots\cdot(b_1a_1^{-1}b_1^{-1}a_1)
& &i=2,\ldots,n\nonumber\\
&\kappa_{\nu,\mu}=m_\mu m_\nu & &1\leq \nu<\mu\leq n\nonumber\\
&\kappa_{\nu, n+2i-1}=a_i^{-1}b_i^{-1}a_i m_\nu & &\nu=1,
\ldots,n,\,i=1,\ldots,g\nonumber\\
&\kappa_{\nu, n+2i}=b_i m_\nu & &\nu=1,\ldots,n,\,i=1,\ldots,g\;.\nonumber
\end{align}
Parametrising the corresponding holonomies
as $\text{Hol}(\gamma)=(u_\gamma,-\Ad^*(u_\gamma^\inv)\bj^\gamma)$ and expressing
them in terms of holonomies of the
generators $m_i,a_j,b_j$, we obtain
\begin{align}
\label{ucurvedef}
&u_{a_i}=u_{A_i}\\
&u_{\delta_i}=u_\ai^\inv u_\bi^\inv u_\ai\nonumber\\
&u_{\alpha_i}=u_\ai^\inv u_\bi^\inv u_\ai u_{B_{i-1}}\nonumber\\
&u_{\epsilon_i}=u_\ai^\inv u_\bi^\inv u_\ai u_{K_{i-1}}\cdots u_{K_1}\nonumber\\&u_{\kappa_{\nu,\mu}}=u_{M_\mu}u_{M_\nu}\nonumber\\
&u_{\kappa_{\nu, n+2i-1}}=u_\ai^\inv u_\bi^\inv u_\ai
u_{M_\nu}\nonumber\\
&u_{\kappa_{\nu,n+2i}}=u_\bi u_{M_\nu}\nonumber\\
\nonumber\\
\label{jcurvedef}
&\bj_{a_i}=\bj_{A_i}\\
&\bj_{\delta_i}=(1-\Ad^*(u_\ai^\inv u_\bi^\inv u_\ai))\bj_\ai-\Ad^*(u_\bi^\inv u_\ai)\bj_\bi\nonumber\\
&\bj_{\alpha_i}=\bj_{B_{i-1}}+(\Ad^*(u_{B_{i-1}})-\Ad^*(u_{\alpha_i}))\bj_\ai-\Ad^*(u_\bi^\inv u_\ai u_{B_{i-1}})\bj_\bi\nonumber\\
&\bj_{\epsilon_i}=\sum_{l=1}^{i-1} \Ad^*(u_{K_{l-1}}\cdots u_{K_1})\bj_{H_l}+(\;\Ad^*(u_{K_{i-1}}\cdots u_{K_1})-\Ad^*(u_\ai^\inv u_\bi^\inv u_\ai u_{K_{l-1}}\cdots u_{K_1})\;)\bj_\ai\nonumber\\
&\qquad\quad- \Ad^*(u_\bi^\inv u_\ai u_{K_{l-1}}\cdots u_{K_1})\;)\bj_\bi\nonumber\\
&\bj_{\kappa_{\nu,\mu}}=\bj_{M_\nu}+\Ad^*(u_{M_\nu})\bj_\mu\nonumber\\
&\bj_{\kappa_{\nu, n+2i-1}}=\bj_{M_\nu}+(\Ad^*(u_{M_\nu})-\Ad^*(u_{\kappa_{\nu,n+2i-1}}))\bj_\ai-\Ad^*(u_\ai u_{\kappa_{\nu,n+2i-1}} )\bj_\bi\nonumber\\
&\bj_{\kappa_{\nu,n+2i}}=\bj_{u_{M_\nu}}+\Ad^*(u_{M_\nu})\bj_\bi,\nonumber
\end{align}
with $u_{K_i}=[u_{B_i},u_{A_i}^\inv]=u_\bi u_\ai^\inv u_\bi^\inv
u_\ai$ and $\bj^{H_i}$ given by \eqref{jothi}.

A Dehn twist around an embedded curve $\gamma$ can be defined by embedding a small
annulus around the curve and twisting its ends by an angle of $2\pi$
as shown in Fig.~4. It induces an outer automorphism of the fundamental group, affecting
only those elements for which all representing curves intersect with
 $\gamma$. If we choose a representing curve that has the smallest possible
 number of intersections with $\gamma$ for each of the generators $m_i$,
 $a_j$, $b_j$ of the
 fundamental group and determine their transformations by
drawing the images of these curves as indicated in Fig.~4,
we obtain explicit formulae for the action of the pure mapping class
group on these generators. We summarise this action in the following table, listing only the generators that do not transform trivially\footnote{Note that the Dehn twist \eqref{dd} is the inverse
  of the twist in \cite{we}.}.

\begin{align}
\label{ad}
a_i: &b_i \mapsto b_ia_i\\
\label{dd}
\delta_i: &a_i \mapsto a_i\delta_i=b_i^{-1}a_i\\
\nonumber\\
\intertext{}
\label{alpha}
\alpha_i: &a_i\mapsto b_i^{-1}a_ib_{i-1}=a_i\alpha_i\\
&b_{i-1}\mapsto
b_{i-1}^{-1}a_i^{-1}b_ia_ib_{i-1}a_i^{-1}b_i^{-1}a_ib_{i-1}=
\alpha_i^{-1}b_{i-1}\alpha_i\nonumber\\
&a_{i-1}\mapsto b_{i-1}^{-1}a_i^{-1}b_ia_ia_{i-1}=
\alpha_i^{-1}a_{i-1}\nonumber\\
\nonumber\\
\nonumber\\
\label{eps}
\epsilon_i: &a_i\mapsto b_i^{-1}a_ik_{i-1}\ldots k_1=a_i\epsilon_i\\
&a_k\mapsto k_1^{-1}\ldots k_{i-1}^{-1}(a_i^{-1}b_ia_i) a_k
(a_i^{-1}b_i^{-1}a_i)k_{i-1}\ldots k_1=\epsilon_i^{-1}
a_k\epsilon_i\nonumber\\
&\qquad\qquad\forall 1\leq k\nonumber
<i\\
&b_k\mapsto k_1^{-1}\ldots k_{i-1}^{-1}(a_i^{-1}b_ia_i) b_k
(a_i^{-1}b_i^{-1}a_i)k_{i-1}\ldots k_1=\epsilon_i^{-1}b_k\epsilon_i\nonumber\\
&\qquad\qquad\forall 1\leq k <i\nonumber\\
\nonumber\\
&\text{where}\; k_j:=b_j a_j^{-1}b_j^{-1}a_j\nonumber\\
\nonumber\\
\nonumber\\
\label{etapp}
\kappa_{\nu,\mu}: &m_\nu\mapsto m_\nu^{-1}m_\mu^{-1}m_\nu m_\mu
m_\nu=\kappa_{\nu,\mu}^{-1}m_\nu\kappa_{\nu,\mu}\\
&m_\mu\mapsto m_\nu^{-1}m_\mu
m_\nu=\kappa_{\nu,\mu}^{-1}m_\mu\kappa_{\nu,\mu}\nonumber\\
&m_\kappa\mapsto  m_\nu^{-1}m_\mu^{-1}m_\nu m_\mu m_\kappa m_\mu^{-1}
m_\nu^{-1} m_\mu m_\nu & &\quad\qquad\qquad\qquad\quad\quad\forall
\nu<\kappa<\mu    \nonumber\\
\nonumber\\
\nonumber\\
\label{etapdelta}
\kappa_{\nu, n+2i-1}: &m_\nu\mapsto
m_\nu^{-1}a_i^{-1}b_ia_im_\nu a_i^{-1}b_i^{-1}a_im_\nu=
\kappa_{\nu,n+2i-1}^{-1}m_\nu\kappa_{\nu, n+2i-1}\\
&a_i\mapsto b_i^{-1}a_im_\nu=a_i\kappa_{\nu,n+2i-1}\nonumber\\
& x_j\mapsto m_\nu^{-1}a_i^{-1}b_ia_im_\nu a_i^{-1}b_i^{-1}a_i
x_ja_i^{-1}b_ia_im_\nu^{-1}a_i^{-1}b_i^{-1}a_i
m_\nu\nonumber\\
&\qquad=\kappa_{\nu,n+2i-1}^{-1}m_\nu\kappa_{\nu,n+2i-1}m_\nu^{-1}x_j m_\nu
\kappa_{\nu,n+2i-1}^{-1}m_\nu^{-1}\kappa_{\nu,n+2i-1}, \nonumber\\
&\qquad\qquad
x_j\in\{m_{\nu+1},\ldots,m_n,a_1,\ldots,b_{i-1}\}\nonumber\\
\nonumber\\
\nonumber\\
\label{etapb}
\kappa_{\nu,n+2i}: &m_\nu\mapsto m_\nu^{-1}b_i^{-1}m_\nu
b_im_\nu=\kappa_{\nu,n+2i}^{-1}m_\nu\kappa_{\nu,n+2i}\\
 &b_i\mapsto
 m_\nu^{-1}b_im_\nu=\kappa_{\nu,n+2i}^{-1}b_i\kappa_{\nu,n+2i}\nonumber\\
&a_i\mapsto m_\nu^{-1}b_i^{-1}a_i b_i^{-1}m_\nu^{-1}b_i
m_\nu\nonumber\\
&\qquad=\kappa_{\nu,n+2i}^{-1}a_ib_i^{-1}m_\nu^{-1}\kappa_{\nu,n+2i}\nonumber\\
&x_j\mapsto
m_\nu^{-1}b_i^{-1}m_\nu b_ix_jb_i^{-1}m_\nu^{-1}b_i
m_\nu\nonumber\\&\qquad=\kappa_{\nu,n+2i}^{-1}m_\nu b_ix_jb_i^{-1}
m_\nu^{-1}\kappa_{\nu,n+2i},\qquad x_j\in\{m_{\nu+1},\ldots,m_n, a_1,\ldots b_{i-1}\} \nonumber.
\end{align}

\vbox{
\vskip .3in
\input epsf
\epsfxsize=12truecm
\centerline{
\epsfbox{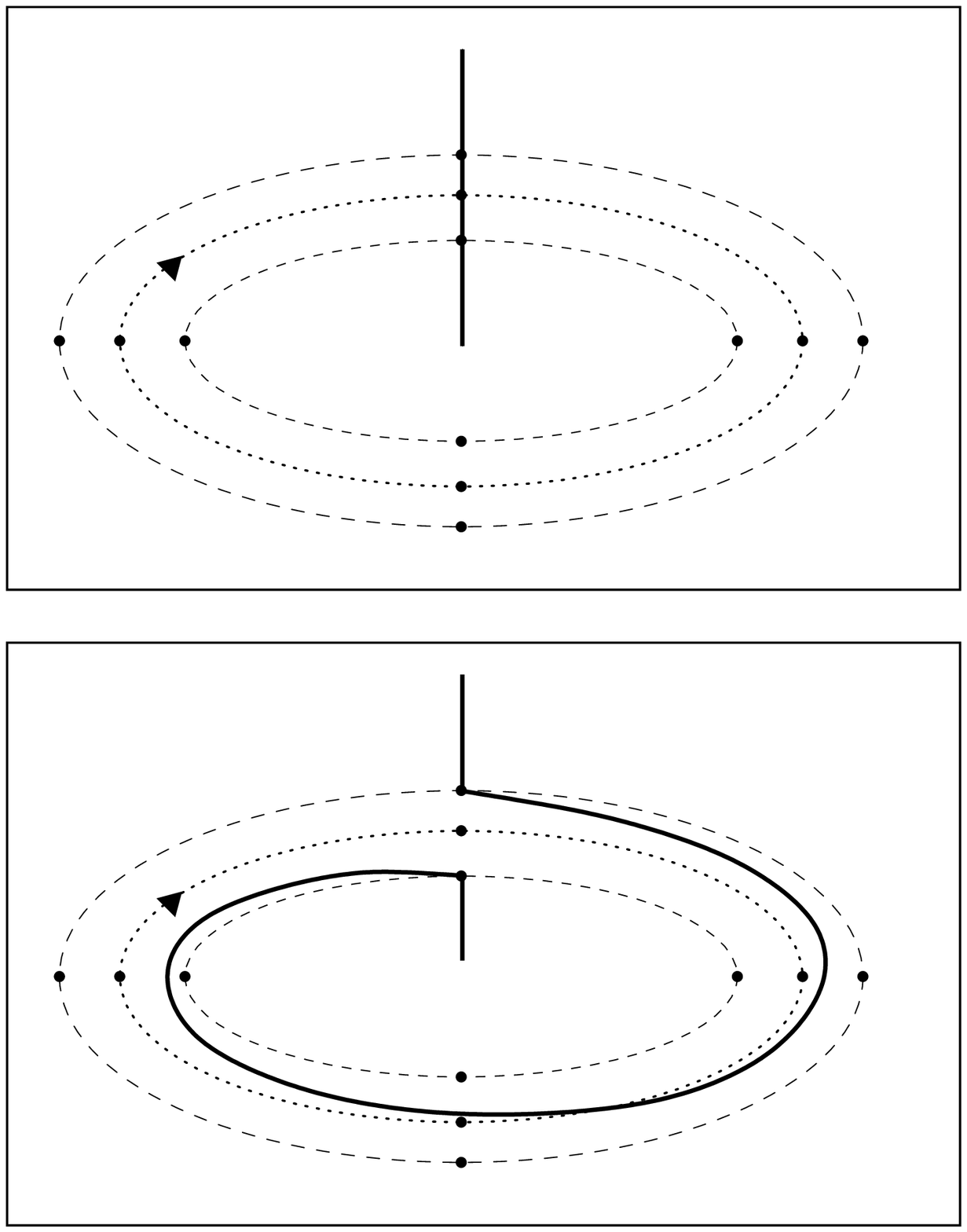}
}
\bigskip
{
\centerline{\bf Fig.~4}
}
\centerline{\footnotesize The effect of a 
 Dehn twist around an oriented loop (dotted
  line)}
\centerline{\footnotesize on  a curve intersecting the loop 
transversally (full line)}
\bigskip}

 A set of generators of the full mapping class group of the surface
 $S_{g,n}\mindee$ is obtained by supplementing this set of generators with
 the  generators $\sigma^i$, $i=1,\ldots,n$ of the braid group. 
The action of these generators on the loops $m_i$ around 
the punctures
is shown in Fig.~5. They leave invariant all generators of the 
fundamental group except $m_i$ and $m_{i+1}$, on which they act according to
\begin{align}
\label{braid}
\sigma^i:
&m_i\mapsto m_{i+1}\\
&m_{i+1}\mapsto m_{i+1}m_im_{i+1}^{-1}\;.\nonumber
\end{align}

\vbox{
\vskip .3in
\input epsf
\epsfxsize=11truecm
\centerline{
\epsfbox{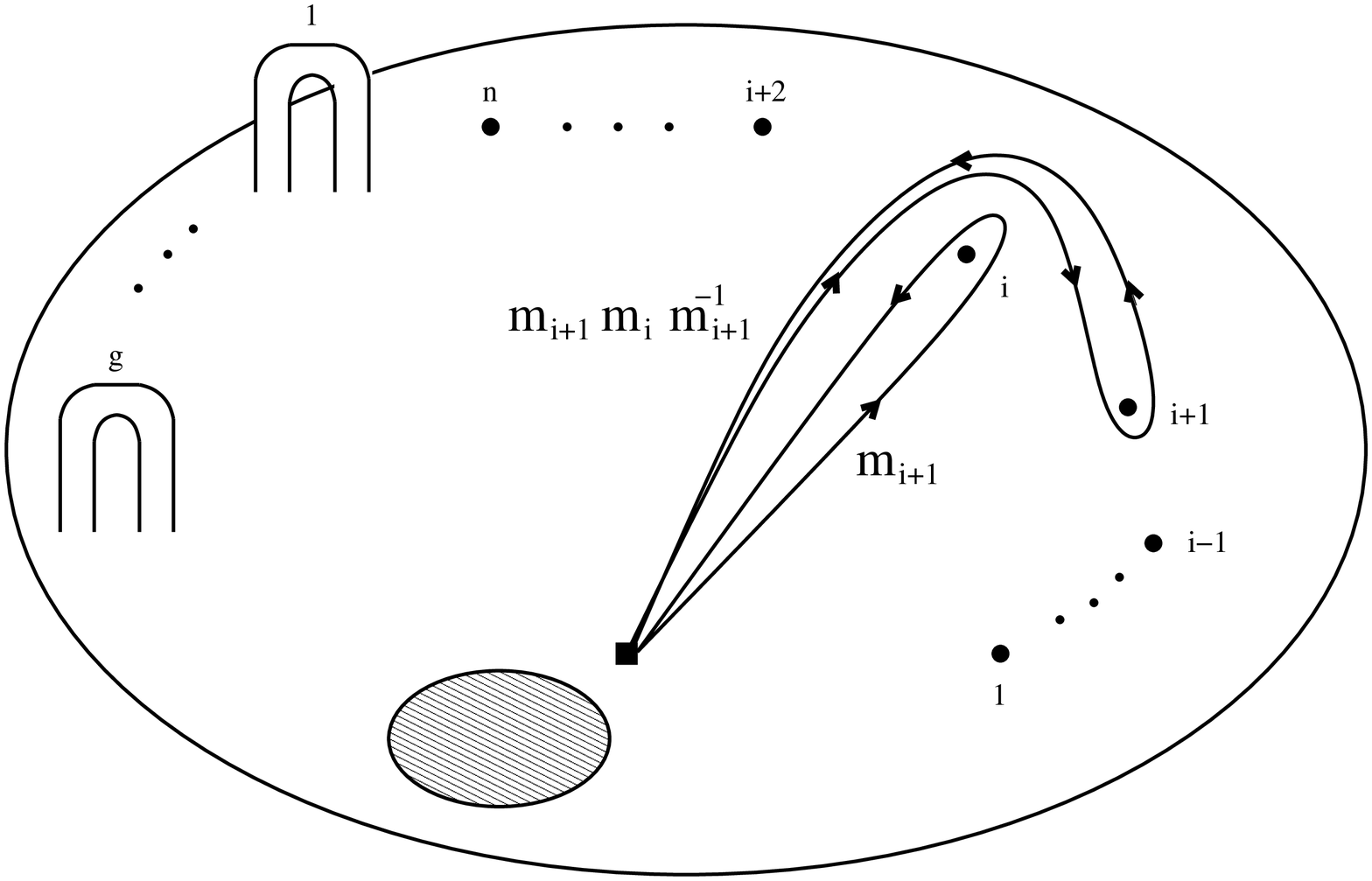}
}
\bigskip
{
\centerline{\bf Fig.~5 }
}
\centerline{\footnotesize The generators of the braid group on the
  surface 
$S_{g,n}\mindee$}
\bigskip}


\begin{thebibliography}{99}

\itemsep=\smallskipamount
\bibitem{AbMars} Abraham, R., Marsden, J.~E.: {Foundations of
    Mechanics}. Cambridge: Perseus Publishing, 1978
\bibitem{AMII} Alekseev, A.~Y., Malkin, A.~Z.: { Symplectic
structure of the moduli space of flat connections on a Riemann
surface}. Commun.~Math.~Phys.{\bf 169},  99-119 (1995)
\bibitem{AGSI} Alekseev, A.~Y., Grosse, H., Schomerus,V.: {
Combinatorial quantization of the Hamiltonian Chern-Simons Theory}.
Commun.~Math.~Phys. {\bf 172}, 317--358 (1995)
\bibitem{AGSII} Alekseev, A.~Y., Grosse, H., Schomerus,V.: {
Combinatorial quantization of the Hamiltonian Chern-Simons Theory
II}. Commun.~Math.~Phys. {\bf 174}, 561--604 (1995)
\bibitem{AS} Alekseev, A.Y., Schomerus, V.: { Representation
theory of Chern-Simons observables}. Duke Math.~Journal {\bf 85},
 447--510 (1996)
\bibitem{AB}
Atiyah, M., Bott, R.: Yang-Mills equations over Riemann surfaces.
{ Phil.~Trans.~R.~Soc.~London} A {\bf 308}, 523 (1982)
\bibitem{BM} Bais, F.~A.,  Muller, N.:
Topological field theory and the quantum double of
  $SU(2)$. { Nucl.~Phys.} {\bf B530},  349--400 (1998)
\bibitem{BMS} Bais, F.~A., Muller, N.~M., Schroers, B.~J.:    
{Quantum  group symmetry and particle scattering in 
(2+1)-dimensional quantum gravity}. Nucl.~Phys.  {\bf B640}, 3--45
(2002)
\bibitem{birmgreen} Birman, J.~S.: {Mapping class groups and their
    relationship to braid groups}. {Comm.~Pure Appl.~Math.} {\bf 22},
    213--38 (1969)
\bibitem{birmorange} Birman, J.~S.: {Braids, links and mapping
    class groups}. {Ann.~of Math.~Studies} {\bf 82}, Princeton: Princeton
    Univ. Press, 1975
\bibitem{BNR} Buffenoir, E., Noui, K., Roche, P.:  Hamiltonian
Quantization of Chern-Simons theory with $SL(2,\CC)$ Group.
{Class.~Quant.~Grav.} {\bf 19}, 4953-5016 (2002)
\bibitem{Carlip}  Carlip, S.:  {Exact quantum scattering in
(2+1)-dimensional gravity}. {Nucl.~Phys. B} {\bf 324},  106--22 (1989)
\bibitem{CP} Chari, V., Pressley, A.: Quantum Groups. Cambridge: Cambridge
  Univ. Press, 1994
\bibitem{FR} Fock, V.~V.~, Rosly, A.~A.: { Poisson structures on
moduli of flat connections on Riemann surfaces and $r$-matrices}. 
ITEP preprint, { 72-92}  (1992) [math.QA/9802054]
\bibitem{Goldman} Goldman, W.~M.:  The symplectic nature of fundamental
groups of surfaces. { Advances in Mathematics} {\bf 54},  200--225
(1984)
\bibitem{Goldman0} Goldman, W.~M.: {Invariant functions on Lie Groups and 
Hamiltonian flows of surface group representations}.
Invent.~Math. {\bf 85}, 263--320 (1986)
\bibitem{geomtop} Ivanov, N.~V.: Mapping Class Groups; in:
  Daverman, R.~J., Sher, R.~B. (Eds.): Handbook of Geometric Topology.
  Amsterdam: Elsevier Science, 2002
\bibitem{kM} Koornwinder, T., Muller, N.~M.: The quantum double
of a (locally) compact group. {Journal of Lie Theory} {\bf 7},
101--120 (1997)
\bibitem{KBM} Koornwinder, T., Muller, N., Bais, F.~A.: Tensor
product representations of the quantum double of a compact group. 
{ Commun.~Math.~Phys} {\bf 198},  157--186 (1998)
\bibitem{MarsRat} Marsden, J.~E., Ratiu, T.~S.: {Introduction to
mechanics and symmetry}. New York: Springer Verlag, 1999
\bibitem{MatPol} Mateev, S., Polyakov, M.: A geometrical presentation
of the surface mapping class group and surgery. Commun. Math. Phys.
{\bf 160}, 537--550 (1994)
\bibitem{we} Meusburger, C., Schroers, B.~J.: Poisson structure and
  symmetry in the Chern-Simons formulation of (2+1)-dimensional
  gravity.
{ Class.~Quant.~Grav.} {\bf 20},  2193--2233 (2003)
\bibitem{we2} Meusburger, C., Schroers, B.~J.: The quantisation of Poisson
    structures arising in Chern-Simons theory with gauge group
    $G\ltimes \mathfrak{g}^*$ , to appear in Advances in Theoretical and 
Mathematical Physics [hep-th/0310218]  
\bibitem{Wajnryb} Wajnryb, B.: A simple presentation of the mapping
class group of an orientable surface. Israel Journal of Mathematics
{\bf 45}, 157--174 (1983)
\bibitem{Witten1} Witten, E.: 2+1 dimensional gravity as an exactly
 soluble system.  Nucl.~Phys. B {\bf 311}, 46--78 (1988) 



\end{thebibliography}
\end{document}